\documentclass{JHEP3}

\usepackage{amsmath,amsfonts,amssymb,graphicx,subfigure,textcomp}
\usepackage[OT1,T1]{fontenc}

\def\be#1#2\ee{\begin{equation}\label{#1}#2\end{equation}}
\def\bea#1#2\eea{\begin{eqnarray}\label{#1}#2\end{eqnarray}}
\def\note#1{\marginpar{\raggedright\if@twoside\ifodd\c@page\raggedleft\fi\fi\sf\scriptsize Note: #1}}
\newcommand{\bZ}{\mathbb{Z}}
\newcommand{\cN}{\mathcal{N}}\newcommand{\cR}{\mathcal{R}}
\newcommand{\A}{A}\newcommand{\B}{B}
\newcommand{\AAA}{AAA}\newcommand{\AAB}{AAB}
\newcommand{\ABA}{ABA}\newcommand{\ABB}{ABB}
\newcommand{\BBA}{BBA}\newcommand{\BBB}{BBB}
\newcommand{\ZZ}{\bZ_2\!\times\!\bZ_2}
\newcommand{\N}{\mathbf{N}}
\newcommand{\Sym}{\mathbf{Sym}}\newcommand{\Anti}{\mathbf{Anti}}
\newcommand{\rmap}{\stackrel{\cR}{\to}}
\newcommand{\lrmap}{\stackrel{\cR}{\leftrightarrow}}
\newcommand{\e}{\varepsilon}\newcommand{\te}{\tilde{\varepsilon}}
\newcommand{\const}{\mathrm{const}}
\newcommand{\rep}[1]{\mathbf{#1}}
\newcommand{\1}{\rep{1}}
\newcommand{\2}{\rep{2}}
\newcommand{\3}{\rep{3}}
\newcommand{\incclipfig}[1]{%
  \includegraphics[width=0.49\linewidth]{#1}}
\newcommand{\fig}[3]{\begin{figure}[t]\begin{center}%
  \includegraphics[width=0.7\linewidth,trim=0mm 15mm 0mm 15mm,clip]{#1}%
  \caption{#3}\label{#2}\end{center}\end{figure}}
\newcommand{\widefig}[3]{\begin{figure}[t]\begin{center}%
  \includegraphics[width=\linewidth]{#1}\caption{#3}\label{#2}%
  \end{center}\end{figure}}
\newcommand{\twofig}[4]{\begin{figure}[t]\begin{center}%
  \subfigure[\label{#3_a}]{\incclipfig{#1}}\hfill%
  \subfigure[\label{#3_b}]{\incclipfig{#2}}%
  \caption{#4}\label{#3}\end{center}\end{figure}}
\newcommand{\fourfig}[6]{\begin{figure}[t]\begin{center}%
  \subfigure[\label{#5_a}]{\incclipfig{#1}}\hfill%
  \subfigure[\label{#5_b}]{\incclipfig{#2}}\\%
  \subfigure[\label{#5_c}]{\incclipfig{#3}}\hfill%
  \subfigure[\label{#5_d}]{\incclipfig{#4}}\\%
  \caption{#6}\label{#5}\end{center}\end{figure}}
\preprint{\small{NIKHEF/2007-006}\\[-1ex]\small{MPP-2007-22}\\[-1ex]\small{LMU-ASC 14/07}\\[-1ex]\small{hep-th/0703011}}
\title{Statistics of intersecting D--brane models on $\mathbf{T^6/\bZ_6}$}
\author{Florian Gmeiner$^{1}$, Dieter L\"ust$^{2,3}$ and Maren Stein$^{3}$\\
~\\
{}$^1$NIKHEF,\\
\phantom{$^1$}Kruislaan 409, 1098 SJ Amsterdam, The Netherlands\\
~\vspace{-1ex}\\
{}$^2$Max--Planck--Institut f\"ur Physik,\\
\phantom{$^2$}F{\"o}hringer Ring 6, D--80805 M{\"u}nchen, Germany\\
~\vspace{-1ex}\\
{}$^3$Arnold--Sommerfeld--Center for Theoretical Physics,\\
\phantom{$^3$}Department f{\"u}r Physik, Ludwig--Maximilians--Universit{\"a}t,\\
\phantom{$^3$}Theresienstra{\ss}e 37, D--80333 M{\"u}nchen, Germany\\
~\vspace{-1ex}\\
E-mail: {\sf fgmeiner@nikhef.nl, luest@mppmu.mpg.de,
         mstein@theorie.physik.uni-muenchen.de}
\bigskip}

\abstract{%
We perform a statistical analysis of supersymmetric intersecting D--brane
models on the type II orientifold $T^6/\bZ_6$.
After providing an analytic proof of the finiteness of the number of possible
solutions in this setup we study the frequency distributions of properties
of the gauge group and the chiral matter content.
In particular we search for models with a standard model gauge group and
discuss their statistical suppression. The results are compared with the
recent studies on $T^6/(\ZZ)$.
The analysis is conducted using a statistical method, based on the choice of
random subsets of the full ensemble of solutions. This method allows to
calculate the total number of models with high precision to~$3\times 10^{28}$.}
 

\begin{document}

%
%
\section{Introduction}\label{intro}
In order to make contact with low energy physics, the quest to find a realistic
MSSM--like string vacuum is one of the most important tasks for string
phenomenology. In the context of type II orientifolds there has been a huge
amount of activity over the last years to obtain a model that resembles the
standard model as closely as possible\footnote{For reviews on this topic see
e.g.~\cite{lu04,bcls05,bkls06}.}.
Since it is believed that there exists a vast landscape of string
vacua containing a huge number of possible solutions~\cite{lls86,do03},
new methods have to be used to analyse this tremendous realm.

Instead of studying individual solutions, it might be better to analyse an
ensemble of models using a statistical approach~\cite{do03}.
With statistical methods, one can try to answer questions about the
distribution of certain properties within the ensemble of solutions.
These distributions might give important insights into the overall shape
of the landscape. On the one hand, they could be a valuable guide for model
building, giving hints where to look for interesting
solutions\footnote{For recent reviews on distributions on the landscape and
counting of flux compactifications, see~\cite{ku06,doka06, ddk07}.}.
Moreover, the issue of correlations of properties within the ensemble of models
is of great importance. Finding correlations implies that it might be possible
to deduce general aspects of the landscape, independent of specific
models.

Dealing with statistics, there are several caveats not to be overlooked.
One of them concerns the finiteness of solutions~\cite{acdo06}.
If the ensemble to be analysed is not finite, the possibility to make clear
statements is greatly diminished, since one has to rely on properties which
appear in a regular pattern.
The same applies for a random sample, which has to be chosen with great
care, in order to make it a representative subset of the full range of
solutions.

In~\cite{bghlw04} and~\cite{gbhlw05} methods to analyse the open string
sector of intersecting brane models have been developed.
In the second paper a survey of models on a $T^6/(\ZZ)$
orbifold was carried out using a computer based approach. This
technique was also used to analyse the statistics of standard--model--like
as well as $SU(5)$ and flipped $SU(5)$ models on the same orbifold in greater
detail~\cite{gm05,gmst06} (for a summary of the results obtained for this
geometry see also~\cite{gm06}). In~\cite{kuwe05} a survey of standard model
vacua including fluxes has been accomplished for this background.
An analytic proof of the finiteness of solutions to the tadpole and
supersymmetry constraints in the case of an $T^6/(\ZZ)$ orbifold has been
given in~\cite{dota06}.
Moreover a statistical analysis of Gepner model orientifolds was performed
in~\cite{dhs04a,dhs04b,adks06}, and aspects of the heterotic landscape were
discussed e.g. in~\cite{di06,le06}.

It is clear that the statistical analysis performed in the articles mentioned
above for the case of the $T^6/(\ZZ)$ orientifold should be repeated for other
background
geometries in order to see if these previous results are somehow generic and
persist, or if they are substantially different for other spaces.
In this article we use similar methods as in the works described above to
analyse a different intersecting brane setup, namely the IIA orientifold with
intersecting D--branes on the $T^6/\bZ_6$ orbifold.
This class of models is also interesting from a phenomenological perspective,
since it has already been shown that one can construct an intersecting brane
model with three generations of quarks and leptons on this space~\cite{hoot04}.

There are many similarities to the $T^6/(\ZZ)$ case,
but we encounter some new aspects as well. In particular, this background
requires fractional branes, coming from the $\bZ_2$--twisted sector of the
orbifold~\cite{ddg97,digo99}. As it turns out, these fractional branes are
essential for the properties of the low energy theory, in particular for the
existence of chiral matter. Moreover, due to the existence of the fractional
branes the number of solutions to the constraining equations is largely
increased compared to the
$T^6/(\ZZ)$ case, and the statistical distributions are also different.

In order to make statistical statements for the full parameter space,
we use a new method of analysis, based on the choice of random subsets of
solutions\footnote{If not further specified in the text, we will use the term
``solution'' in this article to stand for a specific model that fulfills the
tadpole, supersymmetry and K--theory conditions, which will be given
explicitly in Section~\ref{constraints}.}.
As emphasized in~\cite{dile06}, this has to be done very
carefully, in order to obtain results that resemble the actual frequency
distributions as closely as possible, since ``floating correlations'' could
have the unwanted effect that certain observables are functions of the
considered examples.
Fortunately, as we will show in this article, the results
obtained in this way are indeed sufficiently close to the full results to be
trusted. We are confident that this method could also be applied to different
setups and, since it greatly reduces the amount of necessary computations,
might prove useful for subsequent surveys of the landscape.

\subsection{Outline}\label{outline}
This paper is organised as follows. In Section~\ref{geometry} we will recall
the geometric setup of $T^6/\bZ_6$, explain the orbifold and orientifold
projections and describe the space of three--cycles. Section~\ref{constraints}
contains a discussion of the constraining equations from tadpole cancellation,
supersymmetry and K--theory.
In Section~\ref{finite} we give an analytic proof of the finiteness of possible
solutions to the constraining equations. We explain our methods of statistical
analysis in Section~\ref{methods} and present the obtained results on the
distribution of gauge sector observables in Section~\ref{results}.
In particular, we look for the frequency distribution of models with a
standard model gauge group and their chiral matter content.
Finally we summarise our results and give an outlook to further directions of
research.

%
%
\section{Geometry}\label{geometry}
In this section we will review the geometric setup of the $T^6/\bZ_6$
orientifold and possible D--brane configurations. We will use the notation
and conventions of~\cite{hoot04}, to which we refer for more details on the
geometry and explicit derivations of some of the results we use in the
following.

\subsection{Orbifold and orientifold projections}\label{projections}
We assume a factorisation of the $T^6$ into three two--tori, described by
complex coordinates $z^i$, $i=1,2,3$, on which the orbifold group $\bZ_6$ acts
as
\be{eqorbaction}\nonumber
\theta: z^i\mapsto e^{2\pi iv_i}z^i,
\ee
with the shift vector defined as $\vec{v}_i=\frac{1}{6}(1,1,-2)$. There
exists another possible action, often denoted by $\bZ'_6$, with a different
shift vector (for a recent model building approach on $\bZ'_6$
see~\cite{balo06}). We will not consider the $\bZ'_6$ orbifold in this article,
but plan to come back to it in the future~\cite{ghsXX}.

In addition to the orbifold group we introduce an orientifold projection,
consisting of the reflection of worldsheet parity $\Omega$ and an
antiholomorphic involution $\cR$, which we choose to be complex conjugation,
\be{eqraction}
\cR:z^i\mapsto\bar{z}^i.
\ee

In order for the orbifold and orientifold projections to be compatible,
\eqref{eqraction} has to be an automorphism of the $\bZ_6$ lattice. This
allows for only two possible geometries of the three two--tori, denoted by
$\A$ and $\B$. In the case of an $\A$--geometry the torus lattice is given
by the root lattice of $SU(3)$, spanned by
$\{\sqrt{2}, (1+i\sqrt{3})/\sqrt{2})\}$.
The $\B$--geometry, which corresponds to a $D9$--brane with background--flux
in the dual type IIB picture, can be obtained from the $\A$--case by a
rotation of $e^{-i\pi/6}$.

Choosing different geometries for the three two--tori and considering only
those combinations which cannot be obtained by trivially interchanging the
first and second torus, which transform in the same way under $\theta$, we
obtain six different possible setups, denoted in the following by
$\AAA,\AAB,\ABA,\ABB,\BBA$ and $\BBB$.

\widefig{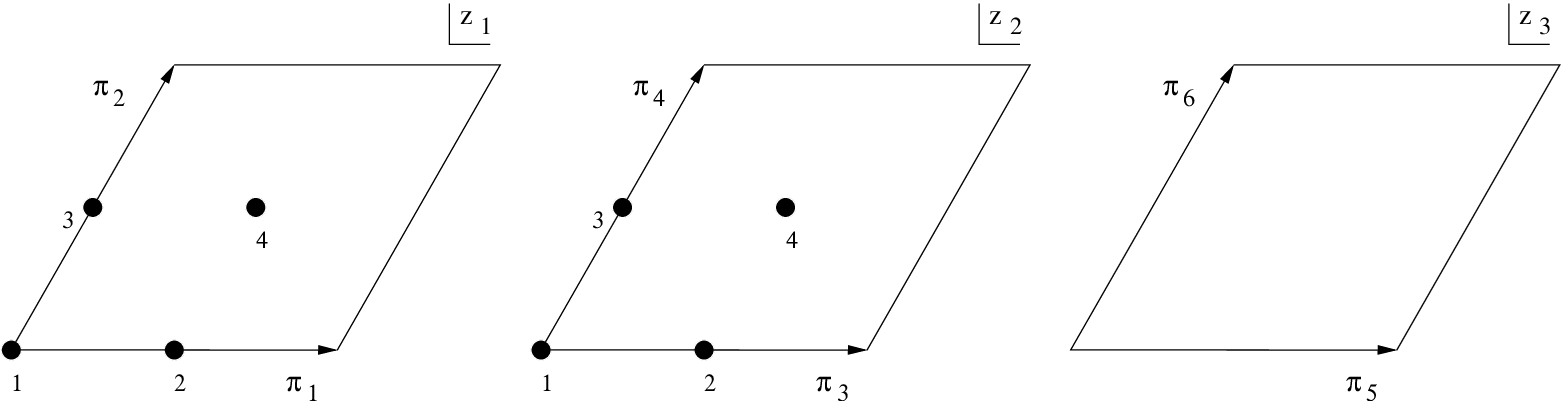}{figt6z6}{The three two--tori of the $T^6/\bZ_6$ orbifold in the
$\AAA$--geometry. The fundamental cycles of the $T^6$ are denoted by $\pi_i$.
The fixed points of $\theta^3$ on the first two $T^2$s, which are relevant for
the definition of exceptional cycles, are marked by dots. The third torus is
invariant under $\theta^3$.}

\subsection{Three--cycles}\label{cycles}
To wrap O6-planes and D6-branes on this geometry, we are interested in the
number of three--cycles, given by the third Betti number $b_3=2(1+h_{2,1})$.
According to~\cite{klra00} we have $h_{2,1}=5$, all coming form the
orbifold--twisted sector. This leads in total to two bulk cycles inherited
from the six--torus and ten exceptional cycles, which wrap a combination of a
one--cycle on $T^3$ and a two--cycle around one of the $\bZ_3$ fixed points.
General three--cycles will be a combination of bulk and exceptional cycles, but
one has to keep in mind that only those combinations are possible in which the
bulk cycle passes through the fixed point in question.

\subsubsection{Bulk cycles}\label{bulkcycles}
The factorisable bulk
cycles can be defined in terms of a basis of fundamental one--cycles on the
three two--tori. For these we use the notation $\pi_{2i-1}, \pi_{2i}$ for
$T^2_i$, $i=1,2,3$, as shown in Figure~\ref{figt6z6}. A basis for the bulk
cycles can be defined as
\bea{eqbasisbulkcycles}\nonumber
  \rho_1=2\left[(1+\theta+\theta^2)\pi_1\otimes\pi_3
                \otimes\pi_5\right],\\
  \rho_2=2\left[(1+\theta+\theta^2)\pi_2\otimes\pi_3
                \otimes\pi_5\right],
\eea
with intersection matrix given by
\be{eqbasisIab}
  I^{(\rho)}_{ij}=\rho_i\circ\rho_j=\begin{pmatrix}0&-2\\2&0\end{pmatrix}.
\ee
Any bulk three--cycle can be expanded using the basis~\eqref{eqbasisbulkcycles}
as
\be{eqbulkcycles}
  \Pi_a=Y_a\rho_1+Z_a\rho_2.
\ee
In terms of the wrapping numbers $n_i,m_i$ of the fundamental one--cycles
$\pi_{2i-1}$ and $\pi_{2i}$ the coefficients $Y_a$ and $Z_a$ read
\bea{eqzycoefficients}\nonumber
  Y_a&=&n_{1,a}n_{2,a}n_{3,a}-m_{1,a}m_{2,a}m_{3,a}-
        \sum_{i}m_{i,a}m_{j,a}n_{k,a},\\\nonumber
  Z_a&=&\sum_{i}m_{i,a}m_{j,a}n_{k,a}+
        \sum_{i}m_{i,a}n_{j,a}n_{k,a},\quad
  i,j,k\in\{1,2,3\} \mathrm{cyclic}.
\eea
From~\eqref{eqbasisIab} and~\eqref{eqbulkcycles} one computes the intersections
between two bulk cycles to be
\be{eqbulkIab}\nonumber
  I_{ab}:=\Pi_a\circ\Pi_b=2(Z_aY_b-Y_aZ_b).
\ee

The action of the involution~\eqref{eqraction} on the fundamental one--cycles
of the two--tori for the two possible geometries $\A$ and $\B$ is given by
\be{eqcycleaction}\nonumber
  \A:\left\{\begin{array}{rcl}\pi_{2i-1}&\rmap&\pi_{2i-1},\\\pi_{2i}&\rmap&
     \pi_{2i-1}-\pi_{2i},\end{array}\right.\qquad
  \B:\left\{\begin{array}{rcl}\pi_{2i-1}&\rmap&\pi_{2i},\\\pi_{2i}&\rmap&
     \pi_{2i-1}.\end{array}\right.
\ee
This leads to the following transformations of the bulk
cycles~\eqref{eqbasisbulkcycles} for the six inequivalent geometries,
\be{eqronbulk}
\begin{array}{l@{\;\;\rho_1\rmap\;}l@{\quad\rho_2\rmap\;}l}
  \AAA: & \rho_1, & \rho_1-\rho_2,\\
  \AAB: & \rho_2, & \rho_1,\\
  \ABA: & \rho_2, & \rho_1,\\
  \ABB: & \rho_2-\rho_1, & \rho_2,\\
  \BBA: & \rho_2-\rho_1, & \rho_2,\\
  \BBB: & -\rho_1, & \rho_2-\rho_1.
\end{array}
\ee
To obtain the cycles wrapped by O6-planes we have to combine two orbits,
invariant under $\Omega\cR\theta^{2k}$ and $\Omega\cR\theta^{2k+1}$,
respectively. For the different geometries we obtain
\be{eqoplanes}
  \Pi_{O6} = \left\{\begin{array}{lll}
    \AAA: 4\rho_1,&\qquad&\ABB: 6\rho_2,\\
    \AAB: 4(\rho_1+\rho_2),&\qquad&\BBA: 4\rho_2,\\
    \ABA: 2(\rho_1+\rho_2),&\qquad&\BBB:4(-\rho_1+2\rho_2).
  \end{array}\right.
\ee

\subsubsection{Exceptional cycles}\label{excycles}
In addition to the three--cycles inherited from the
six--torus we obtain additional, so--called exceptional cycles, which wrap a
product of cycles around the $\theta^3$--orbifold fixed points
(denoted by 1,2,3,4 in Figure~\ref{figt6z6}) and a one--cycle on $T_3$. This
situation is similar to the one that has been encountered in the case of
compactifications on $T^6/\bZ_4$ in~\cite{bgo02}.

We can choose the following basis of exceptional cycles, invariant under the
orbifold projection,
\bea{eqbasisexceptionalcycles}\nonumber
\e_1  &=& (e_{21}-e_{41})\otimes\pi_5+(e_{41}-e_{31})\otimes\pi_6, \\\nonumber
\te_1 &=& (e_{31}-e_{41})\otimes\pi_5+(e_{21}-e_{31})\otimes\pi_6, \\\nonumber
\e_2  &=& (e_{12}-e_{14})\otimes\pi_5+(e_{14}-e_{13})\otimes\pi_6, \\\nonumber
\te_2 &=& (e_{13}-e_{14})\otimes\pi_5+(e_{12}-e_{13})\otimes\pi_6, \\\nonumber
\e_3  &=& (e_{22}-e_{44})\otimes\pi_5+(e_{44}-e_{33})\otimes\pi_6, \\\nonumber
\te_3 &=& (e_{33}-e_{44})\otimes\pi_5+(e_{22}-e_{33})\otimes\pi_6, \\\nonumber
\e_4  &=& (e_{23}-e_{42})\otimes\pi_5+(e_{42}-e_{34})\otimes\pi_6, \\\nonumber
\te_4 &=& (e_{34}-e_{42})\otimes\pi_5+(e_{23}-e_{34})\otimes\pi_6, \\\nonumber
\e_5  &=& (e_{24}-e_{43})\otimes\pi_5+(e_{43}-e_{32})\otimes\pi_6, \\
\te_5 &=& (e_{32}-e_{43})\otimes\pi_5+(e_{24}-e_{32})\otimes\pi_6, 
\eea
where we denoted the two--cycles stuck at the fixed points on $T_1$ and $T_2$
by $e_{ij}$, $i,j=1,\ldots,4$ as shown in Figure~\ref{figt6z6}.
The intersection matrix of these exceptional cycles is given by
\be{eqexceptionalIab}\nonumber
I^{(\e)}_{ij} := \bigoplus_{k=1}^5\begin{pmatrix}
                 \te_k\circ\te_k&\te_k\circ\e_k\\
                 \e_k\circ\te_k&\e_k\circ\e_k\end{pmatrix}
               = \bigoplus_{k=1}^5\begin{pmatrix}0&-2\\2&0\end{pmatrix}.
\ee
Under the action~\eqref{eqraction} of $\cR$ the fixed points on the first two
$T^2$s are mapped into each other as follows,
\be{eqexrmap}\nonumber
  \A:\left\{\begin{array}{rcl}1&\rmap&1,\\2&\rmap&2,\\3&\lrmap&4
     \end{array}\right.\qquad
  \B:\left\{\begin{array}{rcl}1&\rmap&1,\\2&\lrmap&3,\\4&\rmap&4
     \end{array}\right.
\ee

It is possible to combine the bulk cycles~\eqref{eqbasisbulkcycles} and
exceptional cycles~\eqref{eqbasisexceptionalcycles}, including their images
under the orbifold projection, into an unimodular lattice of basic
three--cycles~\cite{hoot04}, which is an important consistency check for
completeness of the symplectic basis.
Since this particular basis is not very convenient for computations, we will
not use it in the following.

\section{Model building constraints}\label{constraints}
In addition to the O6--planes described by~\eqref{eqoplanes}, we introduce
$k$ stacks of D6--branes, wrapping fractional cycles. However, we would like
to obtain supersymmetric models which are stable and free of anomalies.
Therefore the brane configuration has to fulfil several constraining
equations, which we will describe in the following.

\subsection{Tadpole cancellation}\label{tadpoles}
In order to obtain consistent models we have to make sure that the total
charge of the RR seven--forms in the compact space cancels.
This imposes a condition on the cohomology classes of these forms, which can
be reformulated in homology. Denoting the orientifold image of a cycle $\Pi_a$
wrapped by some brane $a$ by $\Pi_{a'}$ it reads
\be{eqRRtadgen}
  \sum_a N_a\left(\Pi_a+\Pi_{a'}\right)-4\Pi_{O6}=0.
\ee
We can split the tadpole condition into two parts containing contributions from
the bulk and exceptional cycles, respectively.
Since the orientifold planes wrap only bulk cycles according
to~\eqref{eqoplanes}, the contributions from D--branes wrapping exceptional
cycles have to cancel among themselves.

Using the basis~\eqref{eqbulkcycles} and the transformation
rules~\eqref{eqronbulk},
we find for the six different geometries the following
conditions\footnote{These conditions can also be derived explicitly by
computing open string amplitudes, see~\cite{hoot04}.} for $k$ bulk branes with
stack sizes $N_a$,
\be{eqtadbulk}\begin{array}{rlcrl}
  \AAA:&\sum_{a=1}^k N_a(2Y_a+Z_a)=16,&\qquad&
  \ABB:&\sum_{a=1}^k N_a(Y_a+2Z_a)=24,\\
  \AAB:&\sum_{a=1}^k N_a(Y_a+Z_a)=16,&\qquad&
  \BBA:&\sum_{a=1}^k N_a(Y_a+2Z_a)=16,\\
  \ABA:&\sum_{a=1}^k N_a(Y_a+Z_a)=8,&\qquad&
  \BBB:&\sum_{a=1}^k N_aZ_a=16.
\end{array}\ee

\subsection{Supersymmetry conditions}\label{susy}
In order to preserve $\cN=1$ supersymmetry, the bulk cycles have to be
calibrated with respect to the same calibration form as the orientifold
planes. In our case of three--cycles, this is the holomorphic three--form and
this means that the cycles have to be special Lagrangian.
Expressed in terms of the expansion coefficients~\eqref{eqbulkcycles} the
conditions are given by
\be{eqsusybulk}\begin{array}{rlcrl}
  \AAA:&Z_a=0,&\qquad&\ABB:&Y_a=0,\\
  \AAB:&Y_a=Z_a,&\qquad&\BBA:&Y_a=0,\\
  \ABA:&Y_a=Z_a,&\qquad&\BBB:&2Y_a=-Z_a.
\end{array}\ee
Since these conditions boil down to the fact that the bulk
branes have to wrap the same cycles as the O6--planes, we obtain the result
that the intersection number between these branes and the orientifold planes
always vanishes,
\be{eqisbbop}
  I_{aO6}=\Pi_a\circ\Pi_{O6}=0.
\ee
To exclude anti--branes from the spectrum, we have to impose one further
condition,
\be{eqantibranes}
  \AAA,\AAB,\ABA:\; Y_a>0,\qquad \ABB,\BBA,\BBB:\; Z_a>0.
\ee
Fractional branes, being a combination of bulk and exceptional cycles,
preserve half of the supersymmetry, if the bulk part
obeys~\eqref{eqisbbop} and~\eqref{eqantibranes}, and the exceptional part
comes from fixed points that are traversed by the bulk cycle.
In total there are 128 different possible combinations of exceptional cycles
for a given bulk cycle. All possible combinations can be found in Tables 23
and 24 of~\cite{hoot04}.

\subsection{K--theory constraints}\label{ktheory}
In addition to the constraints from tadpole cancellation and supersymmetry, we
have to demand that the four--dimensional models are anomaly--free.
Cancellation of local
gauge anomalies is guaranteed by a generalised Green--Schwarz mechanism,
yet there exists the possibility to obtain a global gauge anomaly~\cite{wi82},
which can be deduced from a K--theory analysis.
In the case of our models, this condition requires an even amount of
chiral matter from $Sp(2)$ probe branes, inserted in the geometric
setup~\cite{ur00}.
$Sp(2)$ gauge groups are carried by branes that are invariant under the
orientifold action. Unfortunately this is not the only possible gauge group
for these branes, they can equally well support an $SO(2)$ group.
To differentiate between these two, one has to go beyond the algebraic approach
that suffices to calculate the tadpole, the susy constraints and the chiral
matter content. It is necessary to analyse the open string M\"obius amplitude
for each possible brane.

Fortunately the geometrical setup of the $\bZ_6$ orientifold is such that we
do not have to worry about this issue. In fact, it can be generally proved
that all possible solutions that fulfil the tadpole and susy constraints will
automatically satisfy the stronger condition where all possible
orientifold--invariant probe branes are used.
In this case we obtain the following condition for a model with $k$ stacks of
branes,
\be{eqktheory}
  \sum_{a=1}^kN_a\Pi_a\circ\Pi_{p} \equiv 0 \mod 2,
\ee
and this equation should hold for any probe brane $p$ invariant under the
orientifold map.

Because of this property and the fact that the bulk part of the probe branes
does not intersect with the bulk part of all other branes, several of the
terms in~\eqref{eqktheory} vanish and we can rewrite it as
\be{eqkt2}
  \sum_{i=1}^5\left(\sum_{a=1}^kN_as_a^i\right)r_p^i \equiv 0 \mod 2,
\ee
where the values $s_a^i$ are the coefficients of the cycles of brane $a$ which
are odd under the orientifold projection and the $r_p^i$ parametrise the
cycles of the probe branes which are even under the orientifold map. Note that
we are summing over exactly half of the dimension of the basis of exceptional
three--cycles.
However, not all of the $r_p^i$ are independent, since the probe branes are
bound to be on top of the orientifold planes. An explicit calculation shows
that there exist only eight different possibilities and that the coefficients
$r_p^i$ are always even. Therefore~\eqref{eqkt2} is always fulfilled.

\subsection{Open string spectrum}\label{spectrum}
The massless chiral states arising at the intersection of different stacks
of D--branes and at the intersection of branes with their orientifold
mirrors and the orientifold planes, can be computed from the intersection
numbers. In general a stack of $N$ branes supports a $U(N)$ gauge group, unless
the three--cycle wrapped by this stack lies on top of the orientifold plane.
In this case we are dealing with an $SO(N)$ or $Sp(N)$ group.

To compute the non--chiral spectrum, one has to analyse the open string
amplitudes. In our statistical analysis we will not do so, but concentrate on
the chiral spectrum only. As shown in Table~\ref{tabspec}, we obtain chiral
matter in a bi--fundamental representation at the intersection of two stacks
$a$ and $b$ with $N_a$ and $N_b$ branes, respectively.
In addition there is the possibility for each stack to contribute matter in the
symmetric and antisymmetric representations of the gauge group.
From the discussion in Section~\ref{susy} it follows that the amount of
symmetric and antisymmetric representations will always be the same, since
there can be no contribution from the intersection with the orientifold planes.
Moreover, it is crucial to work with fractional cycles, since all bulk cycles that occur
lie on top of the orientifold plane and do hence not intersect which each other.

\begin{table}[ht]
\begin{center}%
\begin{tabular}{|c|c|}\hline
representations & multiplicity \\\hline
$(\N_a,\overline{\N}_b)$&$\Pi_a\circ\Pi_b$\\
$(\N_a,\N_b)$&$\Pi_a\circ\Pi'_b$ \\
$\Sym_a$&$\frac{1}{2}\left(\Pi_a\circ\Pi_{a'}-\Pi_a\circ\Pi_{O6}\right)$\\
$\Anti_a$&$\frac{1}{2}\left(\Pi_a\circ\Pi_{a'}+\Pi_a\circ\Pi_{O6}\right)$\\
\hline\end{tabular}%
\caption{Multiplicities of the chiral spectrum.}
\label{tabspec}
\end{center}
\end{table}

\subsection{Embedding of the standard model}\label{secsmembed}
Since our final goal is to quantify the number of standard model--like vacua
that can be found in this type of compactifications, we have to chose a way to
realise the gauge group and chiral matter content of the MSSM in terms of
intersecting branes.
In the present work we will consider only one type of embedding, mainly for
two reasons. One is given by external constraints on computational power and
feasibility. The second one lies in the special properties of the orbifold we
are investigating. Since we saw in the previous section that the amount of
symmetric and antisymmetric representations is always equal, several possible
constructions of standard model spectra that use antisymmetric representations
of $SU(2)$ cannot be realised, unless one also allows for chiral matter in the
symmetric representation, which is not desirable from a phenomenological point
of view.

\begin{table}[ht]
\begin{center}%
\begin{tabular}{|c|c|c|c|}\hline
particle & representation & multiplicity \\\hline
$Q_L$ & $(\3,\overline{\2})_{0,0}+(\3,\2)_{0,0}$
      & $I_{ab}+I_{ab'}$\\
$U_R$ & $(\overline{\3},\1)_{-1,0}+(\overline{\3},\1)_{0,-1}$
      & $I_{a'c}+I_{a'd}$\\
$D_R$ & $(\overline{\3},\1)_{1,0}+(\overline{\3},\1)_{0,1}$
      & $I_{a'c'}+I_{a'd'}$\\
$L$   & $(\1,\overline{\2})_{-1,0}+(1,\overline{\2})_{0,-1}
         +(\1,\2)_{-1,0}+(\1,\2)_{0,-1}$
      &  $I_{bc}+I_{bd}+I_{b'c}+I_{b'd}$\\
$E_R$ & $(\1,\1)_{1,1}$ & $I_{cd}$ \\
$N_R$ & $(\1,\1)_{1,-1}$ & $I_{cd'}$ \\\hline
\end{tabular}%
\caption{Realisation of standard model particles with four stacks of branes.
The notation in the second column gives the representation under $SU(3)$ and
$SU(2)$ in brackets with the charges under the $U(1)$s of the third and fourth
stack as subscripts.}
\label{tabsmspec}
\end{center}
\end{table}

The construction we will use in Section~\ref{secsm} for the analysis of the
frequency distribution of standard models is well--known and has been used
in many model building approaches of intersecting branes. It consists of two
stacks of branes ($a$ and $b$) with gauge groups $U(3)$ and $U(2)$, and two
branes ($c$ and $d$) with a $U(1)$ group.
The standard model spectrum is realised through chiral matter
transforming in bifundamental representations of the gauge groups.
The complete spectrum and the assignment to particles is given in
Table~\ref{tabsmspec}.

The hypercharge $Q_Y$ is realised in this construction as a combination of the
$U(1)$ charges $Q_i$, with $i=\{a,b,c,d\}$ of the four branes.
Explicitly it is given by
\be{eqhyper}\nonumber
  Q_Y=\frac{1}{6}Q_a+\frac{1}{2}Q_b+\frac{1}{2}Q_c.
\ee

%
%
\section{Finiteness of solutions}\label{finite}
An important question that we would like to answer before analysing the
four--dimensional models in detail concerns the finiteness of possible
solutions to the constraining equations outlined in Section~\ref{constraints}.
To answer this question, it is sufficient to analyse the solution space of
the system of equations~\eqref{eqtadbulk} and~\eqref{eqsusybulk}.
We do not have to take the analogue expressions for the exceptional cycles into
account, although the set of solutions is greatly enhanced by models containing
exceptional cycles, because the number of possible combinations of these
cycles is always finite (cf. Section~\ref{susy}).
The K--theory constraints will play no r{\^o}le anyway, as has been argued
above.

One important drawback of our approach has to be mentioned here. We cannot
make any statement about the dependence of the number of solutions on the
complex structure moduli\footnote{Concerning this point the present case
differs from the $\ZZ$--case considered in~\cite{dota06}.
On the one hand this is an advantage, because it makes the
proof of finiteness in the $\bZ_6$--case less involved since no free parameters
besides the brane wrapping numbers appear in the constraining equations. On the
other hand we lose a great deal of generality that can only be regained by a
proper analysis of the open string moduli space of the exceptional cycles --
an issue that is beyond the scope of this work.}.
The complex structures of the three two--tori are fixed by the requirement
to be compatible with the orbifold projection. Since $h_{2,1}=5$, we find
ten complex structure moduli in the twisted sector. We do not analyse the
blow up of the orbifold singularities and can therefore not make any statements
about the behaviour of our models away from the orbifold point.
Having said this, we will continue to prove that there is only a finite number
of models at this point in moduli space.

After the susy conditions are fulfilled, we are left with one tadpole
condition for each possible geometry, according to~\eqref{eqtadbulk}.
It will contain one unknown wrapping number ($Y$ or $Z$, depending on
the geometry), which is always positive according to~\eqref{eqantibranes}.
Therefore it follows trivially that the remaining unknown in the tadpole
equations is bounded from above by the orientifold charge, which also depends
on the geometry, but will never be greater than 24.
To proof the finiteness of the number of models, it remains to be shown that
the possible combinations of wrapping numbers $\{n_i,m_i\}$, which make up
$Y$ and $Z$ according to~\eqref{eqzycoefficients}, are always finite.

In the following we will give an explicit proof for the $\AAA$--geometry, the
other five possibilities can be treated analogously.
In order to simplify the discussion and reflect the symmetries of the
problem, we define new variables for the wrapping numbers on the first two
tori, while keeping the wrapping numbers on the third torus explicit.
\bea{eqfinvar}\nonumber
  \alpha &:=& m_1m_2 + n_1m_2 + n_2m_1,\\
  \qquad\beta &:=& n_1n_2+n_1m_2+n_2m_1.
\eea
Exchanging the first two tori, which is a symmetry of the geometric setup,
will leave $\alpha$ and $\beta$ invariant.
In terms of $\alpha,\beta,n_3,m_3$ \eqref{eqsusybulk} reads
\be{eqfin1}
  Z = n_3\alpha + m_3\beta = 0.
\ee
Since we know from~\eqref{eqantibranes} that $Y$ has to be positive, one stack
of branes has to contribute a finite value $0<T<16$ to the tadpole constraint.
This amounts to a second equation,
\be{eqfin2}
  Y = n_3(\beta-\alpha)-m_3\alpha = T.
\ee
To analyse the possibility of an infinite set of solutions to~\eqref{eqfin1}
and~\eqref{eqfin2}, we have to distinguish between the cases
$n_3=0$ and $n_3\neq 0$.

\paragraph{$\mathbf{n_3=0}$:}
Since $n_3$ and $m_3$ cannot vanish simultaneously, we get from~\eqref{eqfin1}
that
\be{eqfin3}
  \beta=n_1(n_2+m_2)+n_2m_1=0.
\ee
and from~\eqref{eqfin2} we obtain
\be{eqfin4}
  -m_3\alpha=-m_3\left(m_2(m_1+n_1)+n_2m_1\right)=T.
\ee
An infinite number of solutions can only exist, if there is an infinite series
of solutions to $\beta=0$ or $\alpha=\const$. Both cases can be treated
analogously, so let us pick one of them and examine $\beta=0$.
Again we analyse two cases, depending on the value of $n_1$. If $n_1=0$,
we get from~\eqref{eqfin3} that $n_2=0$ and~\eqref{eqfin4} reads
$-m_1m_2m_3=T$, which puts bounds on $\{m_i\}$.
If $n_1\neq 0$, we can rewrite~\eqref{eqfin3} as
\be{eqfin6}\nonumber
  m_2=-\frac{n_2}{n_1}\left(n_1+m_1\right).
\ee
Substituting this into~\eqref{eqfin4} leads to
\be{eqfin7}\nonumber
  \frac{m_3n_2}{n_1}\left(n_1^2+m_1^2+n_1m_1\right)=T.
\ee
To obtain an infinite series, the expression in brackets would have to have an
infinite number of solutions. This is not possible, since the term always
defines an ellipse, which supports only a finite number of integer--valued
points.

\paragraph{$\mathbf{n_3\neq 0}$:}
In this case we can rewrite~\eqref{eqfin1} as
\be{eqfin8}
  \alpha=-\frac{m_3}{n_3}\beta.
\ee
Substitution in~\eqref{eqfin2} leads to
\be{eqfin9}
  \left(n_3+m_3+\frac{m_3^2}{n_3}\right)\beta=T.
\ee
The expression in brackets defines again an ellipse and can have only a finite
number of solutions. The remaining possibility would be that there exists an
infinite series to $\beta=n_1n_2+n_1m_2+n_2m_1=\const$.

Let us analyse the different possibilities
for $\{n_1,m_1,n_2,m_2\}$. If $n_1$ or $n_2$ are zero, we can see
immediately that an infinite series is impossible. If $m_1$ vanishes instead,
we obtain $n_1(n_2+m_2)=\const$. If there should exist an infinite series,
$n_2$ and $m_2$ have to be unbounded. Using the definition~\eqref{eqfinvar}
we deduce that $\alpha=n_1m_2$ has would be unbounded as well.
This is only consistent with~\eqref{eqfin8} if $m_3$ would grow beyond all
bounds, which is in contradiction to~\eqref{eqfin9}. The argument can be
repeated analogously for $m_2$ vanishing.

So we are left with the case of $n_1,n_2,m_1,m_2$ all non--vanishing. In this
situation we can rewrite $\beta$ as
\be{eqfin10}\nonumber
  n_1n_2\left(1+\frac{m_2}{n_2}+\frac{m_1}{n_1}\right)=\const,
\ee
which can have only an infinite number of solutions, if $m_2n_1=-m_1n_2$ with
$m_1,m_2$ unbounded. But in this case we find $\alpha$ to be unbounded, which
is not possible with $\beta$ bounded at the same time.
This completes the proof that there can only be a finite number of solutions
to the tadpole and supersymmetry conditions.

%
%
\section{Methods of analysis}\label{methods}
In the following we describe the computational methods we used to obtain
an ensemble of solutions to the tadpole, supersymmetry and K--theory
conditions. The results of the
statistical analysis are based on this explicitly calculated ensemble.

\subsection{Choice of basis}
It turns out that it is convenient to use a different basis of three--cycles
for the computational analysis, because it makes the tadpole
conditions~\eqref{eqRRtadgen} for the bulk cycles and exceptional cycles
more uniform.
The basis consists of $\mathcal{R}$ even cycles $\eta_i$ and $\mathcal{R}$ odd
cycles $\lambda_i$, $i=0,\ldots,5$, which are given in terms of the basis of
bulk cycles~\eqref{eqbasisbulkcycles} and
exceptional cycles~\eqref{eqbasisexceptionalcycles} for the different
geometries as\footnote{Note that in comparison to tables 6 and 7
in~\cite{hoot04} we use a slightly different notation. Due to a sign error the
cycles $\eta_I$ and $\chi_I$ in the notation of that article have to be
exchanged for $I=1,\ldots,5$ (cf. the erratum on p. 33). This we take into
account and moreover we will use $\lambda$ instead of $\chi$.}
\bea{eqrefbaseetalambda}\nonumber
  \vec{\eta}&=&\frac{1}{2}\left\{\begin{array}{rl}
  \AAA: & (\rho_1,            -\e_1+2\te_1,
          -\e_2+2\te_2,       -\e_3+2\te_3,
          \te_4-\e_5+\te_5,   \e_4-\e_5),\\
  \AAB: & (\rho_1+\rho_2,     -\e_1+\te_1,
          -\e_2+\te_2,        -\e_3+\te_3,
          \e_4-\te_5,         -\te_4+\e_5),\\
  \ABA: & (\rho_1+\rho_2,     -\e_1+2\te_1,
          2\e_2-\te_2,        -\e_5+2\te_5,
          \te_3-\e_4+\te+4,   \e_3-\e_4),\\
  \ABB: & (\rho_2,            -\e_1+\te_1,
          \e_2,               -\e_5+\te_5,
          \e_3-\te_4,         -\te_3+\e_4),\\
  \BBA: & (\rho_2,            2\e_1-\te_1,
          2\e_2-\te_2,        2\e_3-\te_3,
          \te_4-\te_5,        \e_4-\te_4+\e_5),\\
  \BBB: & (-\rho_1+2\rho_2,   \e_1,
          \e_2,               \e_3,
          -\e_4+\te_4-\te_5,  \e_4+\e_5),
  \end{array}\right.\\\nonumber
  \mbox{and}&&\\\nonumber
  \vec{\lambda}&=&\frac{1}{2}\left\{\begin{array}{rl}
  \AAA: & (-\rho_1+2\rho_2,   \e_1,
          \e_2,               \e_3,
          \e_4+\e_5,          -\te_4-\e_5+\te_5),\\
  \AAB: & (-\rho_1+\rho_2,    \e_1+\te_1,
          \e_2+\te_2,         \e_3+\te_3,
          -\te_4-\e_5,        -\e_4-\te_5),\\
  \ABA: & (-\rho_1+\rho_2,    \e_1,
          -\te_2,             \e_5,
          \e_3+\e_4,          -\te_3-\e_4+\te_4),\\
  \ABB: & (-2\rho_1+\rho_2,   \e_1+\te_1,
          \e_2-2\te_2,        \e_5+\te_5,
          -\te_3+\e_4,        -\e_3-\te_4),\\
  \BBA: & (-2\rho_1+\rho_2,   -\te_1,
          -\te_2,             -\te_3,
          \e_4-\e_5+\te_5,    -\te_4-\te_5),\\
  \BBB: & (-\rho_1,           \e_1-2\te_1,
          \e_2-2\te_2,        \e_3-2\te_3,
          \e_4-\e_5,          -\te_4+\e_5-\te_5).
  \end{array}\right.
\eea
The expansion of a three--cycle in terms of this basis reads
\be{eqrefcycle}\nonumber
  \Pi_a=\vec{r}\cdot\vec{\eta}+\vec{s}\cdot\vec{\lambda}
       =\sum_{i=0}^5 \left(r_a^i \eta_i + s_a^i\lambda_i\right),
\ee
with expansion coefficients $r^i, s^i$, $i=0\ldots 5$. The tadpole equations
are given by
\be{eqreftad}
\sum_a N_a \vec{r}_a =4 \vec{r}_{O6},
\ee
The zeroth entry of $\vec{r}_{O6}$ can be read off from \eqref{eqtadbulk},
while all others have to vanish, since the orientifold planes do not contribute
to the tadpole equations of the exceptional cycles.

In terms of this new basis the intersection between two stacks of branes $a$
and $b$ defined by cycles $\Pi_a$ and $\Pi_b$ reads
\be{eqisrs}
  I_{ab}=\Pi_a\circ\Pi_b
        =\frac{1}{2}\left( \vec{s}_a\cdot\vec{r}_b
                          -\vec{r}_a\cdot\vec{s}_b\right).
\ee

\subsection{Algorithm}\label{secalg}
To obtain a large number of models that fulfil the constraining equations,
we used several computers to generate the solutions, which were subsequently
stored in a database for later analysis. A priori no constraints have been
imposed on the models besides being consistent solutions to the
tadpole and supersymmetry conditions.

As mentioned before, the model building constraints described in
Section~\ref{constraints} can be treated separately for bulk and exceptional
cycles. 
The first part of the computer program we use, which searches for pure bulk
configurations, employs the partition algorithm used in~\cite{gbhlw05} to find
all possible realisations of the left hand side of equation~\eqref{eqreftad}.
Subsequently it runs through a certain range of pairwise coprime wrapping
numbers searching for groups $(n_i,m_i),\medspace i=1,2,3$ that yield the
desired $r^0$ values. Care has to be taken to avoid multiple counting
of cycles which are identified under the orbifold or orientifold action.
Explicitly, two of the wrapping numbers are restricted to be always $> 0$ and
the wrapping numbers on the third torus, $(n_3,m_3)$ have been chosen to be
both odd. In this way no double counting of solutions which are related by
a geometric symmetry of the problem will occur.
Subsequently the program checks the bulk supersymmetry
conditions~\eqref{eqsusybulk} and~\eqref{eqantibranes}, which amount to
$r_a>0,\medspace s_a^0=0$ in the notation introduced above.
Finally one finds configurations of bulk cycles, which fulfil all consistency
conditions, by combining the results of the previous steps.

According to tables 23 and 24 in~\cite{hoot04} $128$ exceptional cycles, which
already satisfy the supersymmetry conditions, arise for one single bulk cycle.
The second part of our program runs through all $128^k$ possible combinations
of exceptional cycles for a bulk configuration with $k$ stacks and checks the
exceptional tadpole conditions explicitly.
Unfortunately there is no way to exclude part of these $128^k$ combinations a
priori and we have no choice but to compute every single one of them in order
to perform a complete analysis. As a consequence the time necessary for the
computation scales exponentially with the number of stacks and could reach the
realm of years or even decades. In the following we will thus only present full
statistics for models with a low number of stacks. For
configurations with a higher number of stacks we randomly select a 
fraction of the $128^k$ possible combinations.
As we will argue in the next section, these randomly chosen subsets can
be trusted to resemble the full statistical distributions and are therefore
sufficient to make statements about frequency distributions of gauge group
properties and chiral matter content.

%
%
\section{Results}\label{results}

In the following we present the results of a statistical analysis of the
ensemble of solutions to the tadpole, supersymmetry and K--theory constraints,
which have been computed as outlined in the last section.

As already mentioned before, a full analysis of all possible models is as yet
impossible. This comes from the simple fact that the total number of solutions
is of the order $10^{28}$, as we are going to show in the following, and an
explicit computation of every single solution is beyond reach of contemporary
computer technology.
Therefore we used the technique of choosing random subsets of possible
solutions which in turn were analysed in detail.
As it turns out this method is perfectly sufficient for a statistical analysis.

After a more detailed explanation of this random method, we
discuss the total number of solutions. Then we turn to discuss
frequency distributions of various properties of the models, in particular the
gauge group factors, the total rank and the chiral matter content.
Finally we look for solutions that realise the gauge group of
the standard model, discuss their suppression within the set of all solutions
and the properties of the hidden sector gauge group.

Along the way we compare the results with an analysis of $\ZZ$ models.
We only cite the relevant results here, a summary of the statistical analysis
that has been done in that case can be found in~\cite{gm06}.

\subsection{Choosing random subsets}\label{secrand}
The most time--consuming part of computing full solutions is
given by adding exceptional cycles to configurations of bulk cycles that
already fulfil the tadpole condition.
As explained in Section~\ref{secalg}, the bulk solutions are obtained
using a fast partition algorithm, while for the exceptional part there is no
other way then to run through all $128^k$ possible combinations and check if
they fulfil the constrains. This algorithm clearly scales exponential with the
number of stacks $k$, such that a complete survey of models with more then
three stacks is not feasible.

\fig{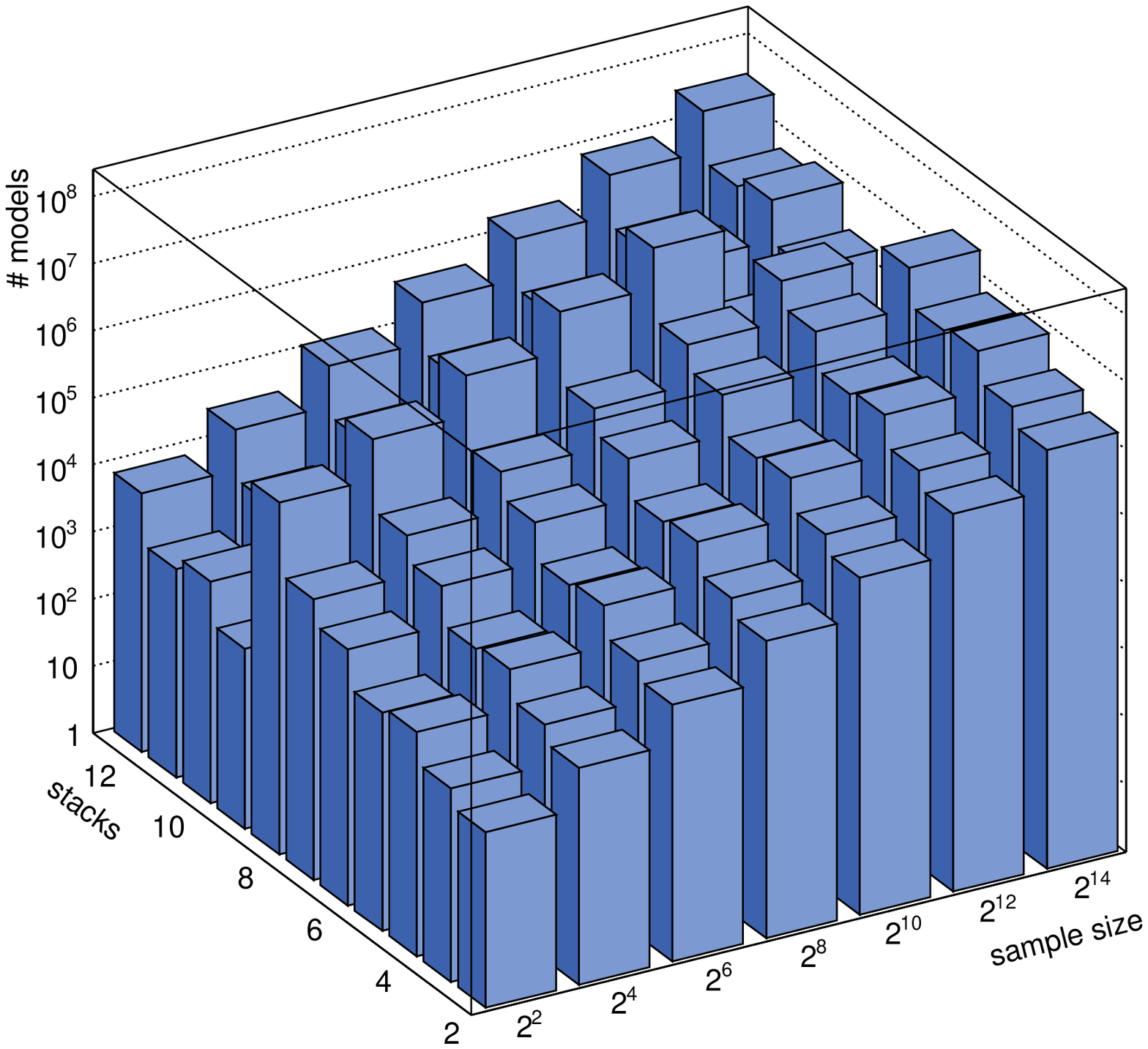}{figlego}{The number of solutions for different numbers of
stacks and sizes of the random sample.}

Nevertheless, as we will explain shortly, we are able to derive quite
robust statistical statements about the full set of solutions.
To do so, we apply a procedure to obtain random subsets of the
$128^k$ possible ways to add exceptional cycles to a bulk solution. If the
total number of solutions is large enough,
it is possible to assume a linear dependence between the size of the sample
$s$ and the number of solutions $n(s)$.
Moreover, the gradient can be used to compute the total number of solutions
$n_{tot}(k)$ for a given stack size, if we assume that this number scales with
$128^k$.

To summarise, we assume that the following equation holds approximately,
\be{eqfit}
  n_{tot}\approx\frac{n(s)}{s}128^k.
\ee
In Figure~\ref{figlego} the number of solutions for different numbers of stacks
and sizes of random samples is shown. Although not clearly visible in this
three dimensional plot, the number of solutions grows indeed linearly with the
size of the random sample.
The accuracy of the linear fit increases with the number of stacks.
According to~\eqref{eqfit}, the slope of the logarithmic plot gives the
average number of full solutions per
bulk configuration, which varies between $10^2$ for the two--stack models and
$4.3\times 10^4$ in the case of models with eight stacks.

\begin{table}[ht]
\begin{center}%
\begin{tabular}{|r|r|r|r|}\hline
stacks & exact & estimate & error\\\hline
$2$ & $1.7068\times 10^6$ & $1.7079\times 10^6$ & $<7\times 10^{-4}$ \\
$3$ & $3.9816\times 10^8$ & $3.9818\times 10^8$ & $<6\times 10^{-5}$ \\\hline
\end{tabular}%
\caption{Exact number of solutions and estimated values for models with two
and three stacks of branes and the relative error of the estimate.}
\label{tabnumsol}
\end{center}
\end{table}

Using the exact results in the two-- and three--stack case we can compare the
total number of solutions with the estimated result from the random procedure.
The results of this comparison are shown in Table~\ref{tabnumsol}.
It turns out that the estimate
is correct up to an error smaller then 0.7\textperthousand~in the case of
models with two stacks and even an order of magnitude less in the case of three
stacks.
Although we cannot completely rule out that something dramatically different
happens for models with a larger number of stacks, this seems very unlikely.
Our results rather suggest that on the contrary one might conjecture that
the estimate gets better for larger stack size $k$.
This can be justified given that the deviation from linearity in the scaling
gets smaller for larger $k$.

It can therefore be expected that the results obtained using the random method
are sufficient for a statistical analysis and that we are allowed to
extrapolate the frequency
distributions obtained for a random sample to the full set of solutions
using the relation~\eqref{eqfit}.
However, it should be emphasised that a good approximation of the number of
solutions is not enough to obtain an accurate description of the properties of
the models.
Therefore we always perform a check for each distribution against the models
with two and three stacks to see if the frequency distributions of the complete
solution and the extrapolated distributions from the random samples do agree.
In particular for properties of the gauge group we expect the method to work
very well, since the gauge group factors depend on the bulk configuration
only.

\subsection{Total number of solutions}\label{sectotnum}
In order to make statistical statements about the probability of certain
properties of solutions, it is certainly important to know about how many
solutions we are talking.
In Figure~\ref{fignumsol} the number of solutions depending on the number
of stacks is given. The left figure shows the number of
solutions to the bulk equations alone, not including exceptional cycles,
while the right figure contains the full result of consistent models.
The minimum number of stacks is two in both cases, while the maximum is twelve,
which can be deduced immediately from the tadpole equations~\eqref{eqreftad}.
Remember that all variables are positive and the maximum value of
the right hand side is $24$, while the wrapping number on the left hand side
is always a multiple of two.

\twofig{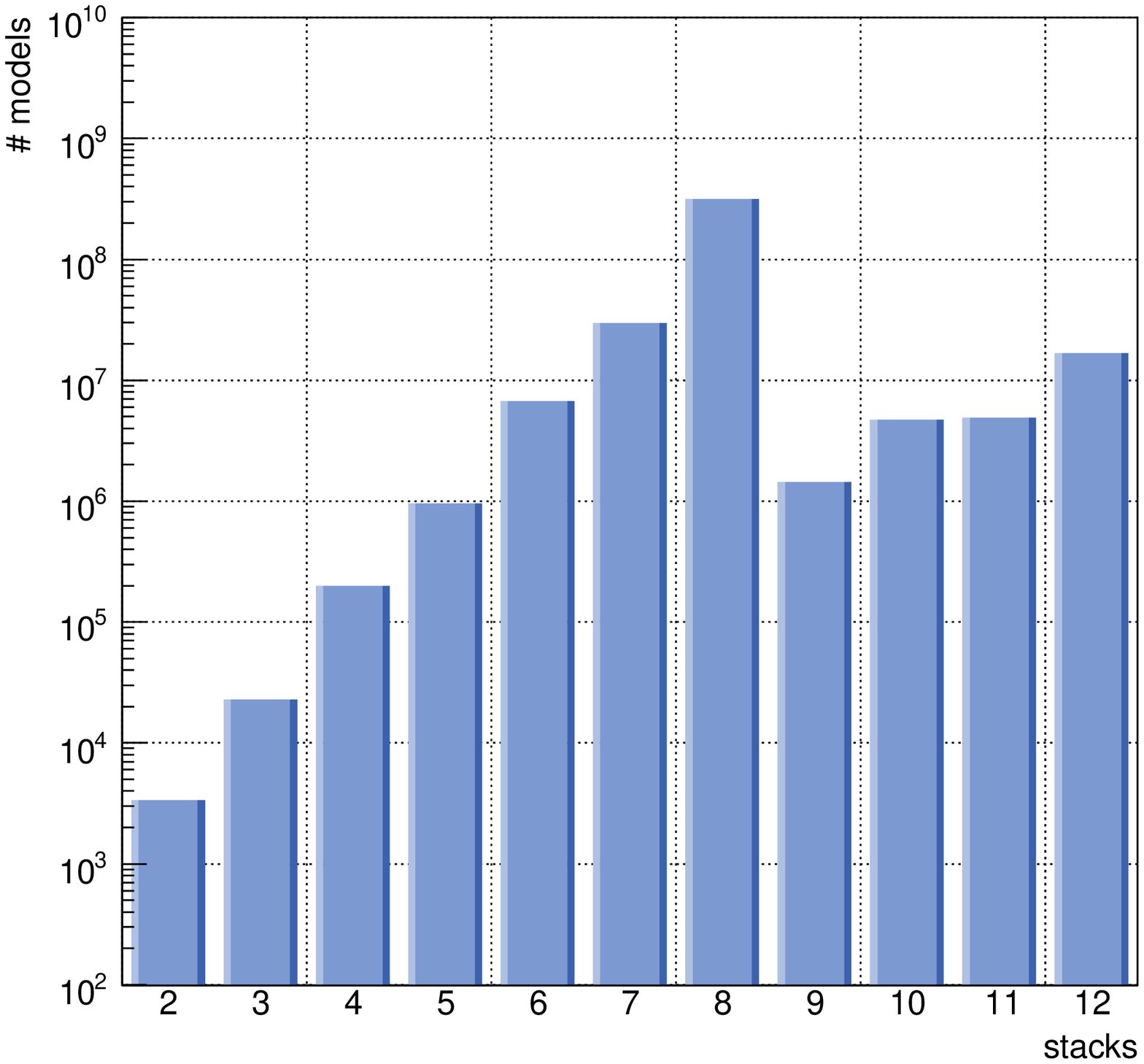}{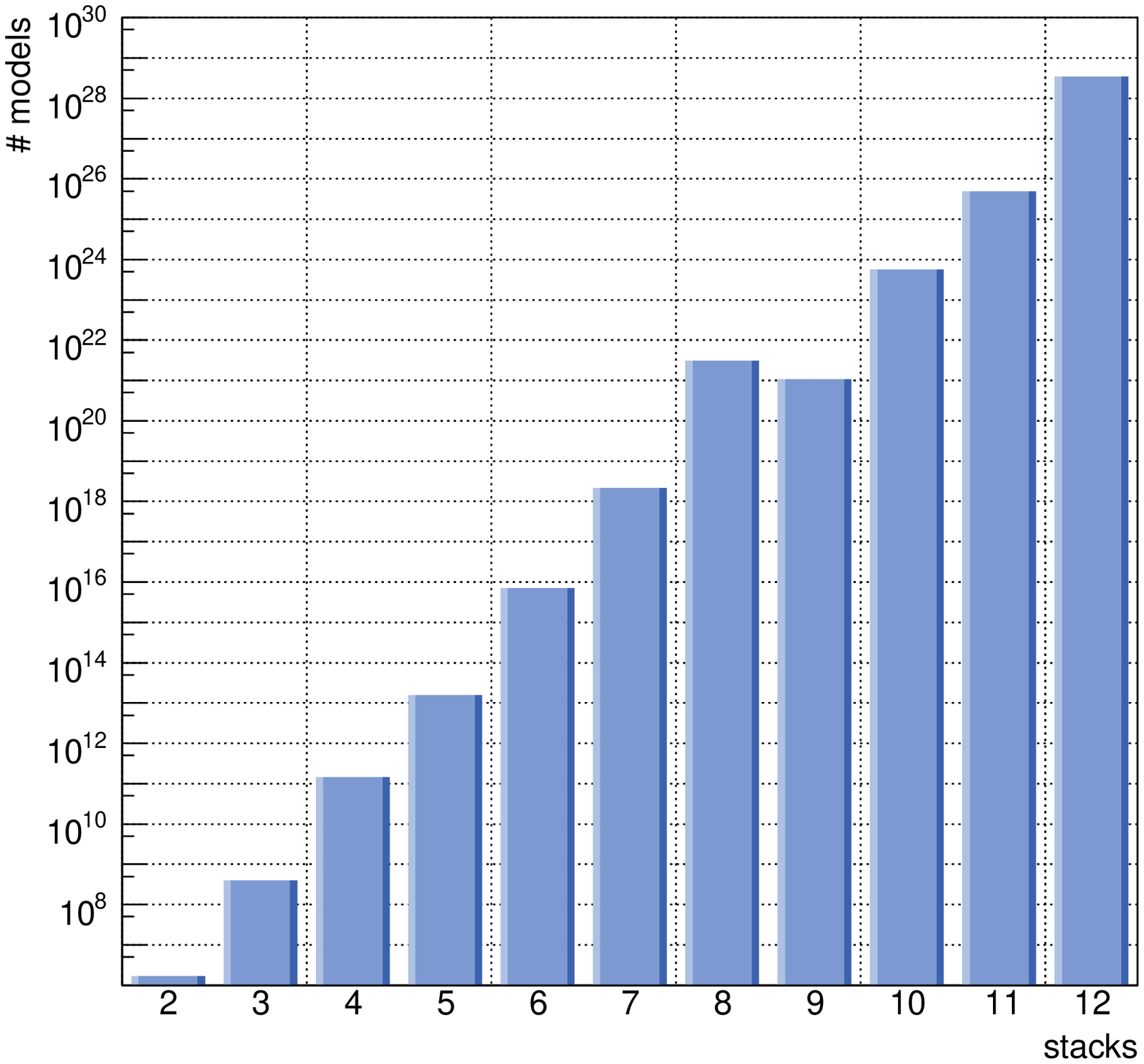}{fignumsol}{Logarithmic plot of the total number
of solutions to the tadpole equations. The left plot (a) shows only bulk
solutions, while the right one (b) show the full set of solutions, including
exceptional cycles.}

Let us begin with an analysis of supersymmetric solutions to the bulk part of
the tadpole conditions~\eqref{eqtadbulk} alone. Note that these configurations
are just an intermediate step to a full solution, since we need to include
exceptional cycles to obtain consistent models.
Nevertheless it is an interesting question to ask how many solutions of the
bulk equations exist, since this gives an overview of the number of candidate
solutions to the full tadpole and susy constraints. As explained above, we
will have to consider $128^k$ possibilities of configurations of exceptional
cycles for each bulk solution.

As one can deduce from Figure~\ref{fignumsol_a}, the maximum number of
solutions of possible bulk cycles is obtained for models with $8$ stacks.
In principle one would assume that the number of possible configurations grows
dramatically with the number of stacks, since naively the number of models
with $k$ stacks should be proportional to the number of factorisations of
integer partitions of length $k$.
However, the negative contribution of the orientifold planes to the tadpole
equation is different for the six possible geometries. For the $\AAA$, $\AAB$,
$\BBA$ and $\BBB$ cases we get a total contribution of $16$, while in the
$\AAB$ and $\ABB$ cases we obtain $8$ and $24$, respectively. Keeping in mind
that there is a factor of two on the left hand side of $\eqref{eqtadbulk}$
and that all brane contributions are positive, one finds that the condition
for models with $\AAB$ geometry can only be fulfilled if the number of stacks
is smaller then five. In the case of $\AAA$, $\AAB$, $\BBA$ and $\BBB$ models
with a maximum of eight stacks are possible. This explains the relatively
small contributions for models with more then eight stacks.

After completing the models with exceptional cycles, the picture changes quite
a bit. This is due to the aforementioned fact that there are in principle
$128^k$ possible configurations of exceptional cycles for each bulk configuration.
Not all of them are consistent, in the sense that they fulfil the full
tadpole equations~\eqref{eqreftad}, but as we have shown in
Section~\ref{secrand} the total number of solutions scales precisely with this
number, multiplied by a coefficient of order $10^2$ to $10^4$. This explains
the domination of models with twelve stacks, that can be seen in
Figure~\ref{fignumsol_b}.
As we will see in the following, this dominance of models with large stack
numbers has a large impact on the statistical distributions.

Using the randomly generated solutions for all possible numbers of stacks we
can compute the total number of models to be $3.43\times 10^{28} \pm 1\%$.
Since the linearity of the growth of solutions increases with large numbers
of stacks, we can estimate the error in this calculation to be smaller then
the relative error calculated explicitly for the two stack models in the last
section.

\twofig{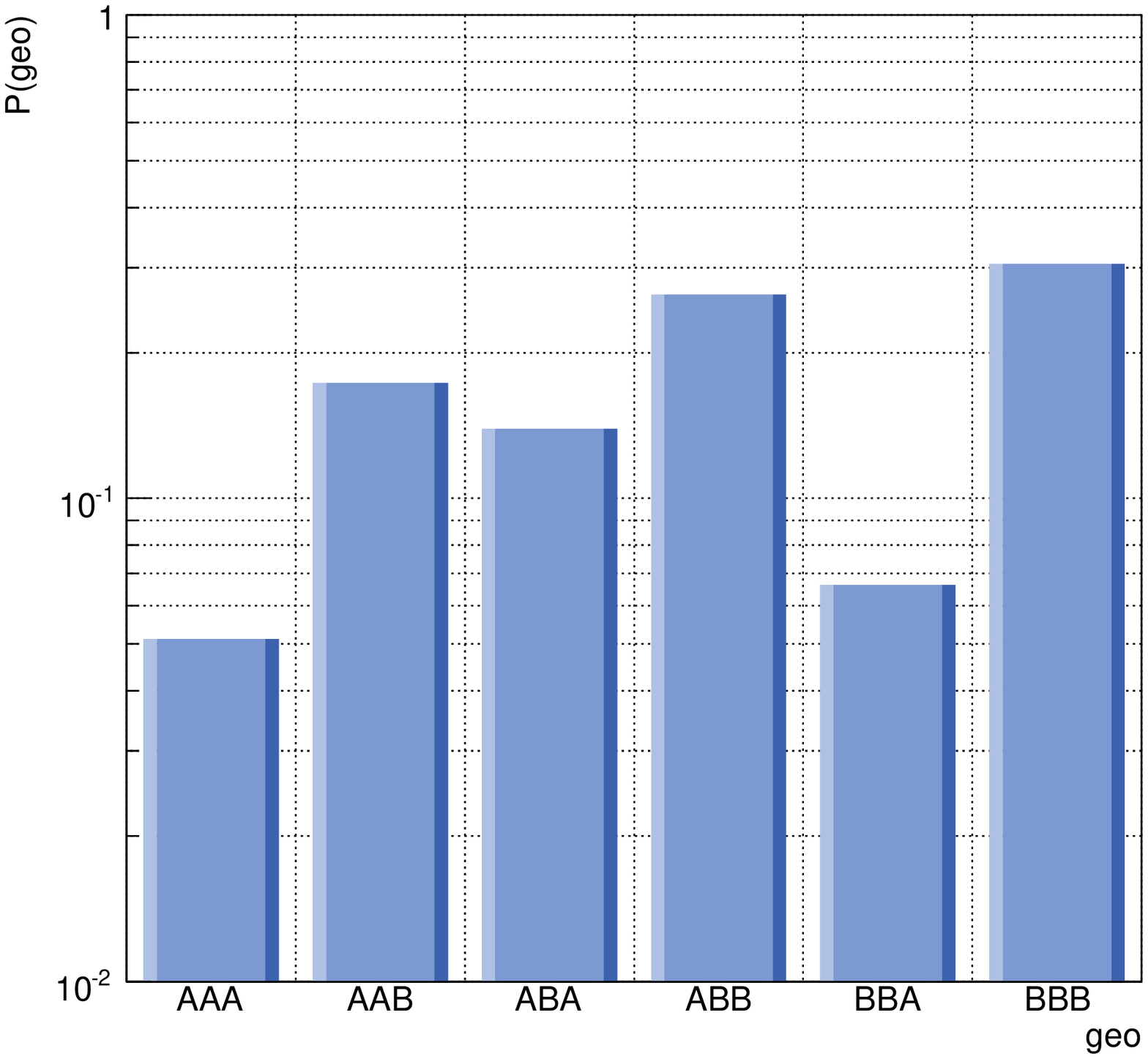}{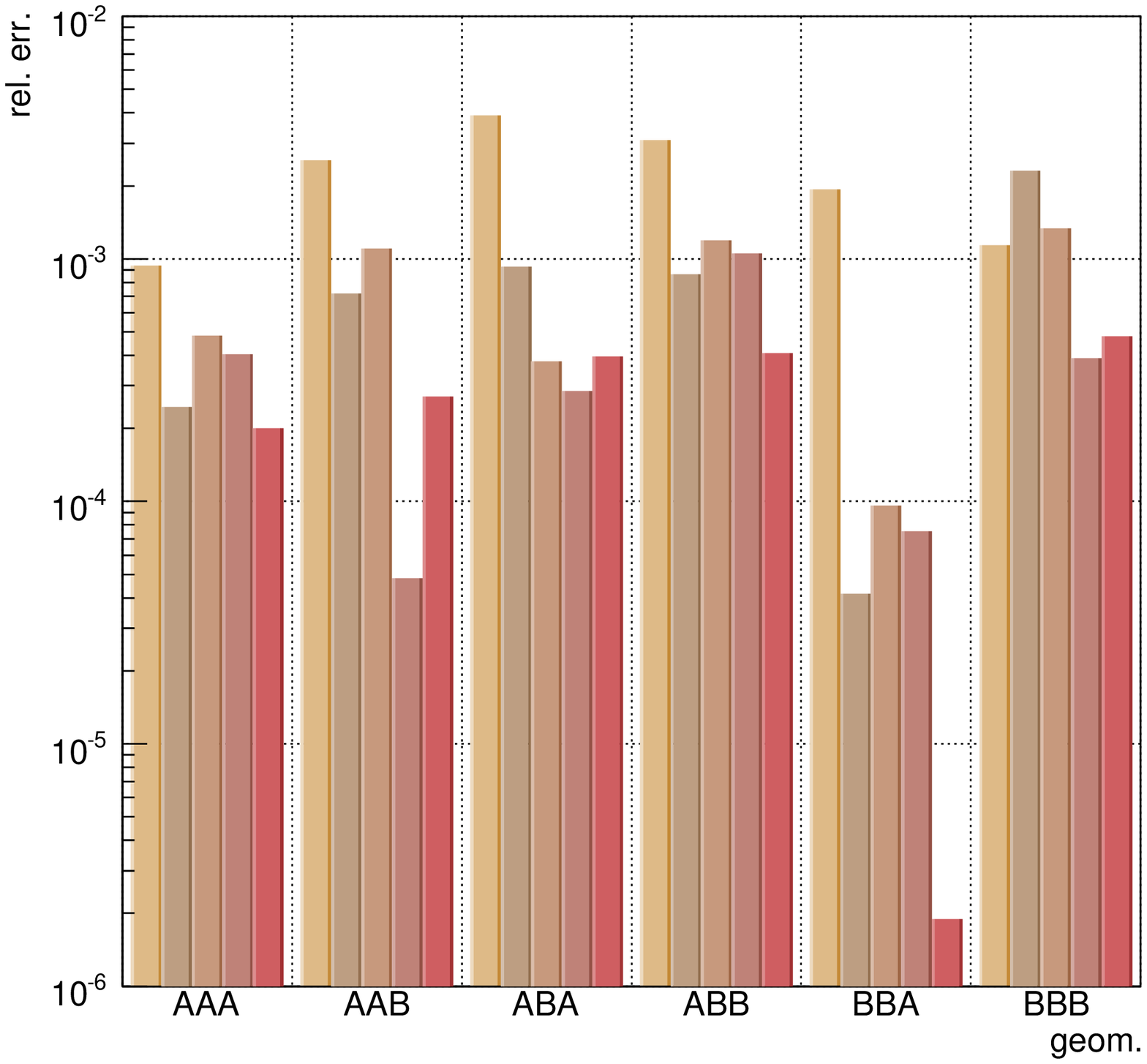}{fignumgeo}{Relative contributions of the
different geometries to the full set of solutions for models with three
stacks. The right figure~(b) shows the relative error between the
random solutions for different stack sizes and the full set of solutions.
The stacks sizes are 4, 16, 64, 256, 1024, 4096 and 16384
(from left to right).}

To complete the picture, we analyse the individual contributions of the
different geometries. This will also serve as a test of the random method
that we used to obtain the statistical distributions. As can be seen in
Figure~\ref{fignumgeo_a}, the largest contribution comes from the $\BBB$
geometry. Concerning the relative error we make using the random method, it
is found to be sufficiently small. As shown in Figure~\ref{fignumgeo_b},
already at a random sample size of $64$ out of $2^{21}$ combinations of
exceptional cycles, we obtain an error smaller then 1\%.

\subsection{Gauge groups}
We consider two properties of the gauge group of the models, which
consists of a product of $U(N)$, $SO(2N)$ and $Sp(2N)$ groups.
Firstly we analyse the distribution of the total rank, defined as
\be{eqtotrk}
  r := \sum_{a=1}^k N_a.
\ee
In a second step we discuss the probability to find one brane with a gauge
group of rank $N$. Both properties are obviously important to
classify models which resemble the standard model.

\subsubsection{Rank distribution}\label{sectotrank}
The frequency distribution of the total rank, see Figure~\ref{figgauge_a},
grows exponentially and reaches a maximum at rank $12$.
This behaviour can be explained by the dominance of models with twelve stacks
of branes. The exponential scaling of the total rank is directly related to
the exponential scaling of the total number of solutions, because these are
dominated by models with an $U(1)$ gauge group. This follows from the
solutions to the tadpole equations, which are given as factorisations of
partitions of the orientifold charge. The factor one is not only the number
with the highest abundance in integer partitions, but it is in fact the only
possible gauge group in models with twelve stacks of branes, as follows
directly from the positivity of all variables in the tadpole equation.

This rank distribution has to be considered with some caution however, since it
includes all possible $U(1)$ factors of the spectrum. It is well known that
some of the $U(1)$s will acquire a mass in the effective theory. This could
happen through a generalised Green--Schwarz mechanism that compensates a
mixed gauge anomaly involving the $U(1)$ in question, but more general cases
are possible. It would be of course very interesting to study the rank
distribution that one obtains after subtracting the massive $U(1)$ factors,
but for the full set of models the necessary computations are not feasible.
In the analysis of models that contain the gauge group of the standard model
in Section~\ref{secsm}, we will make sure that at least one massless $U(1)_Y$
(the hypercharge) exists, but a quantitative statement about the number of
massive $U(1)$s in the general case is beyond the means of our approach.

\twofig{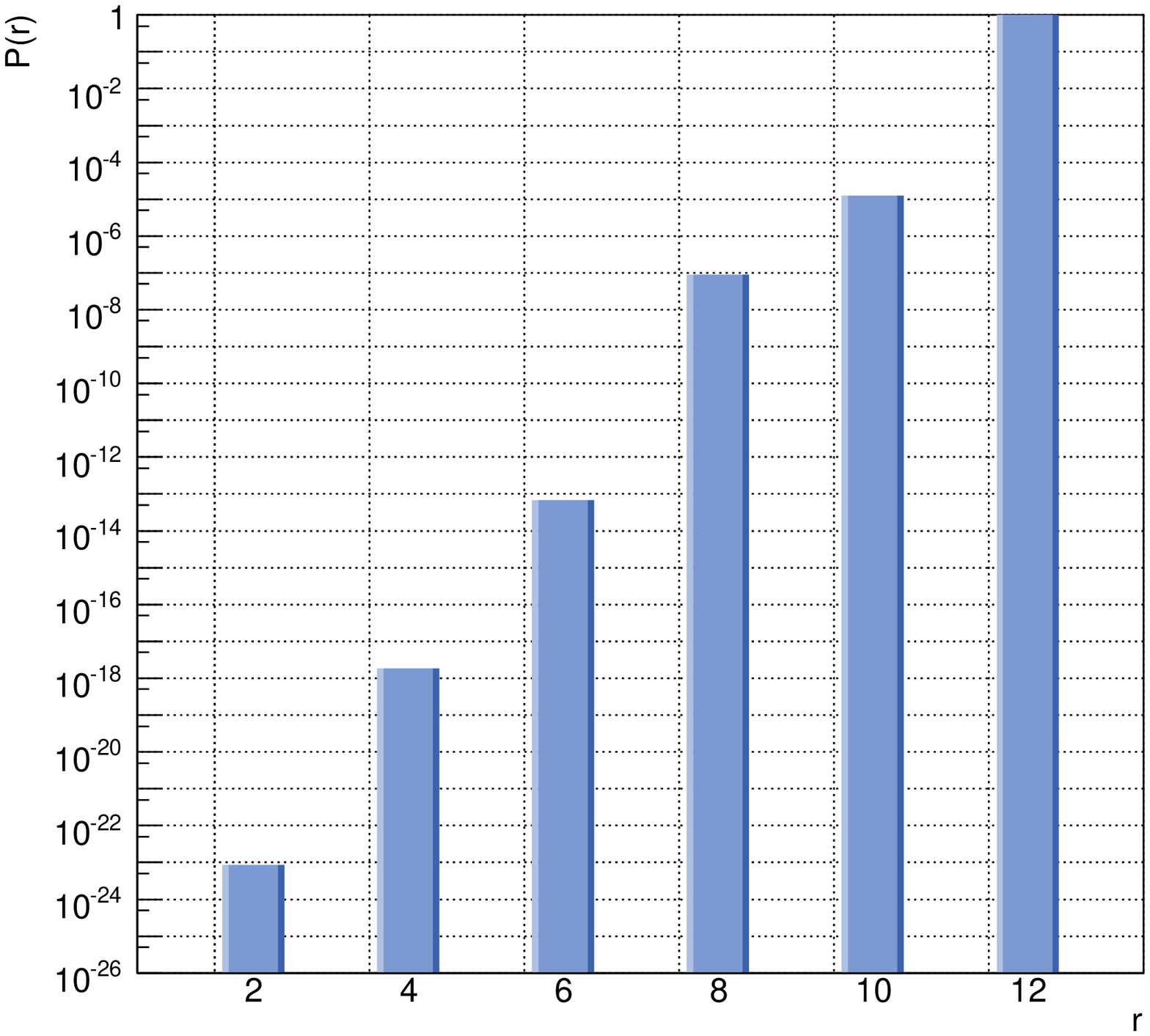}{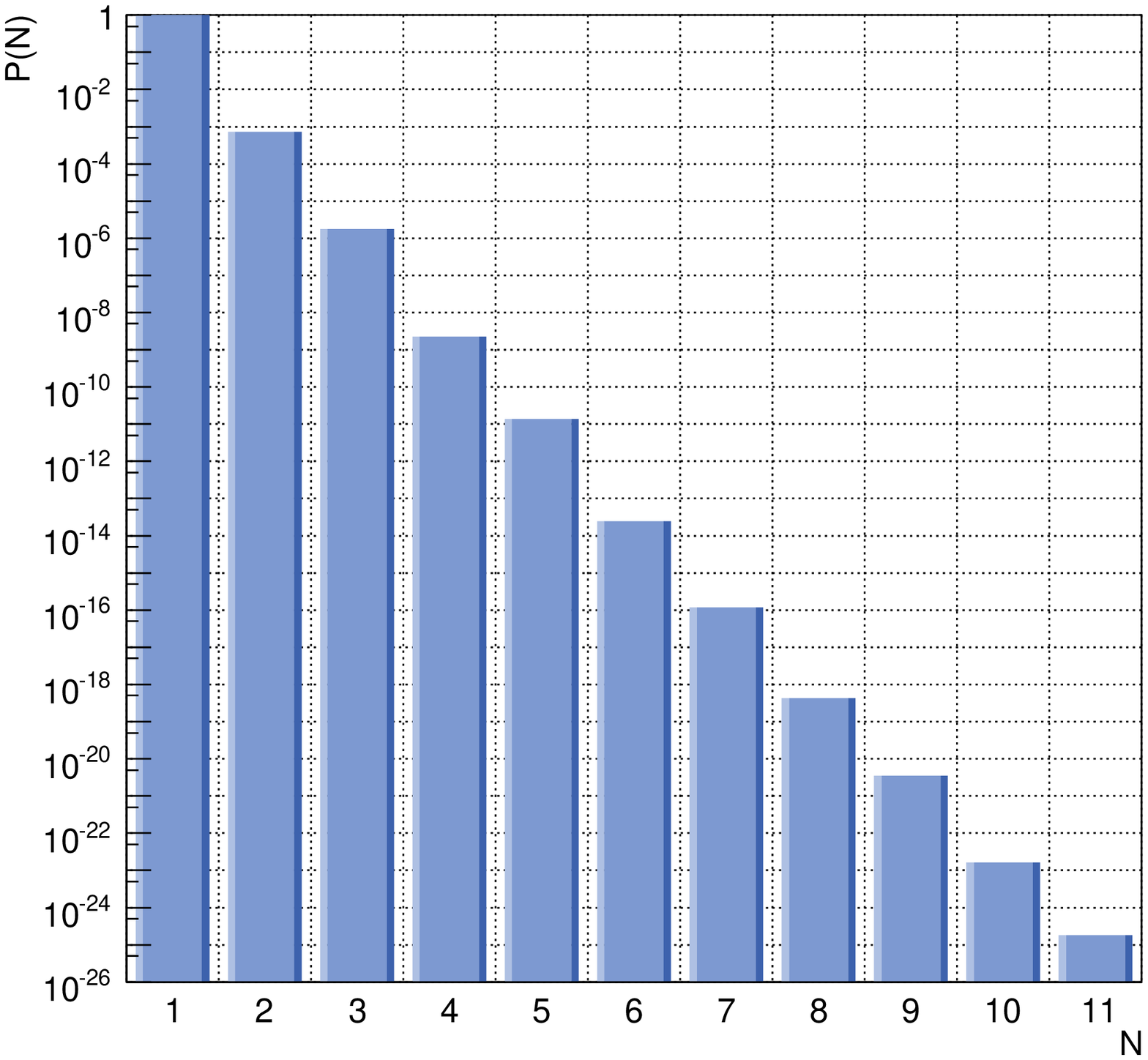}{figgauge}{Frequency distributions of (a)~the total
rank~$r$ and (b)~the probability to find a gauge group of rank~$N$.}

One striking fact of the rank distribution still has to be explained:
There are only solutions with even rank. This is a consequence of the specific
$\bZ_6$ geometry and different from other orbifold models, as for example the
$\ZZ$ models we already mentioned.
To show why this is always the case, we have to take a closer look at the
tadpole equation~\eqref{eqreftad}. The right hand side is always a multiple
of $4$, depending on the geometry. Therefore we have to have
\be{eqevrk1}
  \sum_{a\in A}S_a\equiv 0 \mod 4,\qquad\mbox{with}\quad
    S_a:=N_aY_a,\quad A:=\{1,\ldots,k\}.
\ee
We split the sum over $S_a$ into two parts, consisting of only even
and only odd values:
\be{eqevrk2}\nonumber
  \sum_{a\in A}S_a=\sum_{a\in O}S_a^{(odd)}+\sum_{a\in E}S_a^{(even)},\qquad
    \mbox{with}\quad O\cup E=A,\quad O\cap E=\emptyset.
\ee
The equivalence~\eqref{eqevrk1} can only be fulfilled, if there is an even
number of $S_a^{odd}$. Writing the total rank~\eqref{eqtotrk} as
\be{eqevrk3}
  r=\sum_{a\in A}N_a=\sum_{a\in O}N_a+\sum_{a\in E}N_a,
\ee
we get that the first part of this sum is even. Here we used that all branes
in the set $O$ have to obey $N_a\equiv 1\mod 2$, for
$S_a^{(odd)}\equiv 1\mod 2$. For the second part of~\eqref{eqevrk3} to be even,
it is enough to show that $Y_a$ is always odd. This can be done by writing
the wrapping number in terms of the fundamental torus wrapping numbers, similar
to what we did in Section~\ref{finite}. From
\be{eqevrk4}
  Y=n_2(\alpha-\beta)-m_2(\beta)\quad,\quad Z=n_2\beta+m_2\alpha=0
\ee
and the constraints $(n_2,m_2)\equiv(1,1)\mod 2$, explained in
Section~\ref{methods}, we obtain from the second equation in~\eqref{eqevrk4}
that $\alpha\equiv\beta\mod 2$ and therefore $Y\equiv1\mod 2$.
This completes the proof.

\subsubsection{Gauge group factors}
In Figure~\ref{figgauge_b} the probability to find a gauge group of rank~$N$ is
shown. For the reason explained in the last paragraph, namely the abundance of
$U(1)$ gauge factors, the probability to find one brane with gauge group of
rank one is almost~100\%. The distribution falls off exponentially for larger
$N$, which is again due to the exponential scaling of the number of solutions
with the number of stacks.

\twofig{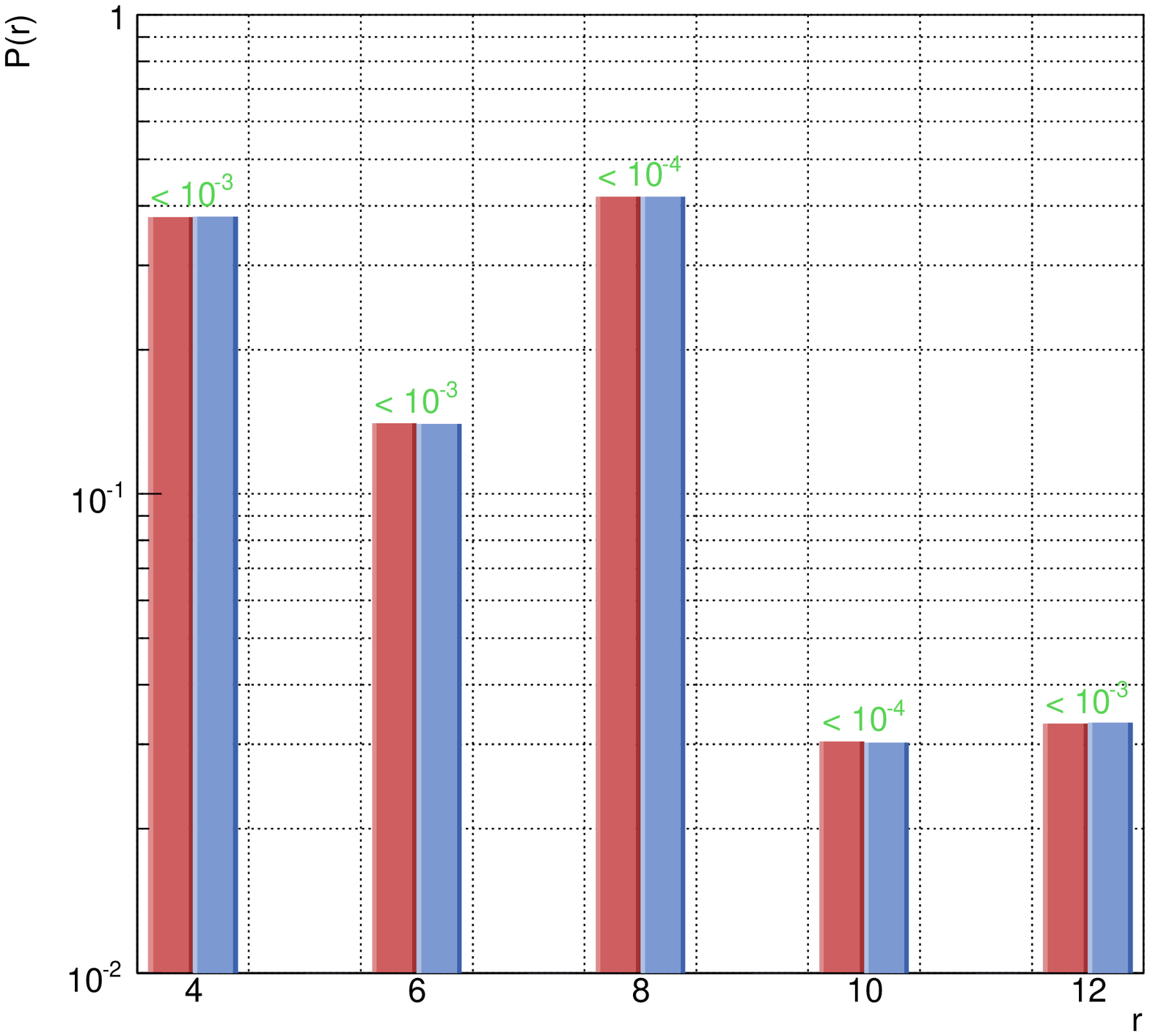}{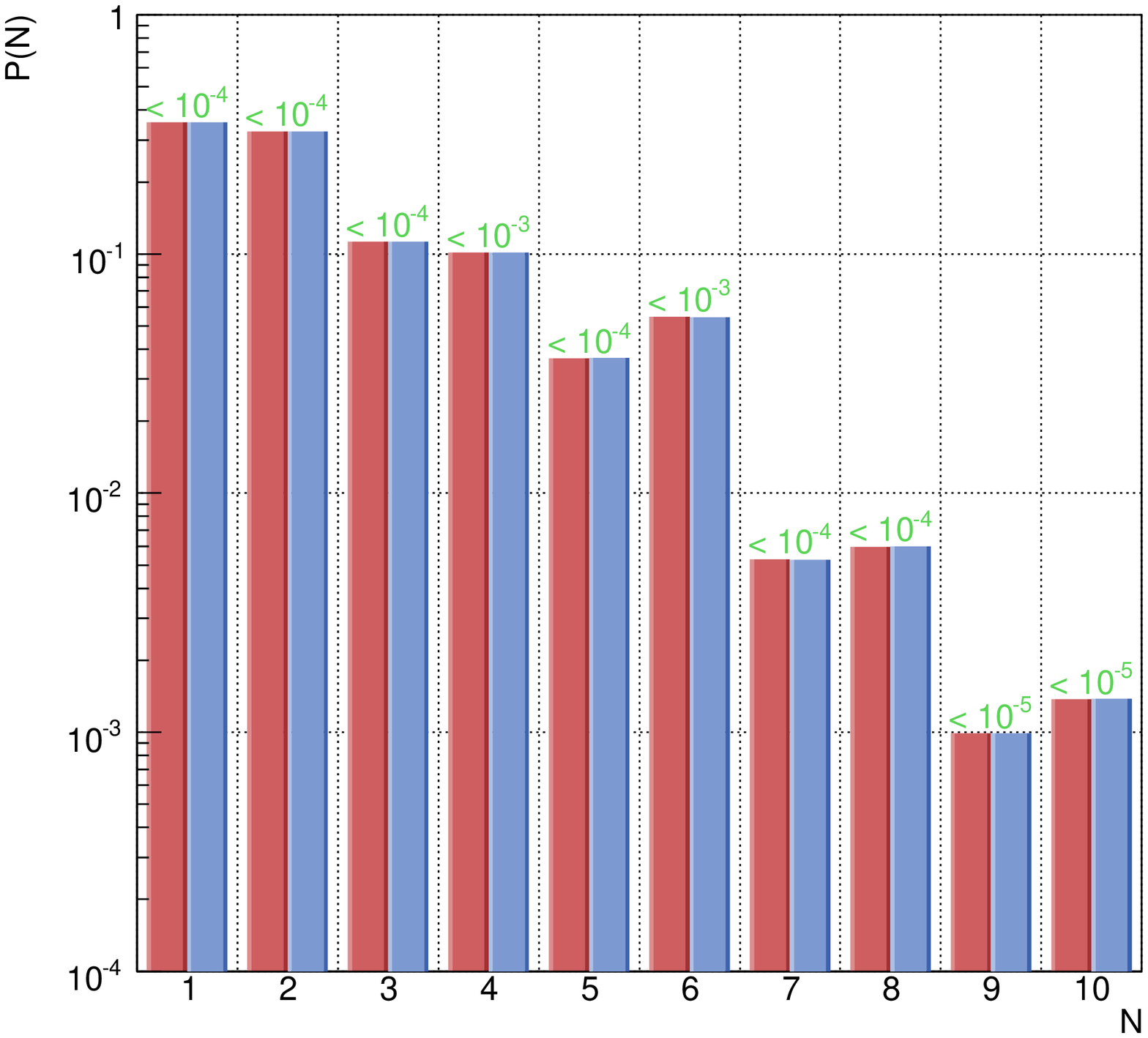}{figgaugecomp}{Comparison of the results for the
distribution of~(a) the total rank $r$ and (b)~the probability to find a
rank~$N$ gauge factor for models with three stacks of branes. The full result
is given by the red bars on the left, while the solutions obtained using a
random set of $2^{14}$ exceptional cycles are shown as blue bars on the right.
An upper bound on the relative error is given by the value above each bar.}

As in the case of the total number of solutions, we have obtained the
distributions using an extrapolation of results from random subsets. To check
the validity of this approach, we compare with the full set of models in
the case of three stacks of branes.
The result is shown in Figure~\ref{figgaugecomp}. For both cases, the total
rank distribution as well as the probability distribution of single gauge
factors, we obtain very accurate results. The relative error is
always smaller then~1\textperthousand in both cases.

\subsection{Mean chirality}
To understand on a qualitative level how many of the solutions are chiral, we
analyse the ``mean chirality'' of the set of solutions. To do so, we define
the mean chirality to be the average of chiral representations in each model.
This definition is identical to the one used in the statistical analysis of
$\ZZ$ orbifold models (cf. Section 3.2.2 of~\cite{gm06}).
For a model with $k$ stacks we define the mean chirality $\chi$ as
\be{eqmeanchi}
  \chi := \frac{2}{k(k+1)}\sum_{\substack{a,b=0\\a<b}}^k
          \left|I_{a'b}-I_{ab}\right|
        = \frac{2}{k(k+1)}\sum_{\substack{a,b=0\\a<b}}^k
          \left|\vec{s}_a\cdot\vec{r}_b\right|,
\ee
where we used the definition of the intersection $I_{ab}$ in terms of the
$\eta,\lambda$--basis~\eqref{eqisrs}.

\fourfig{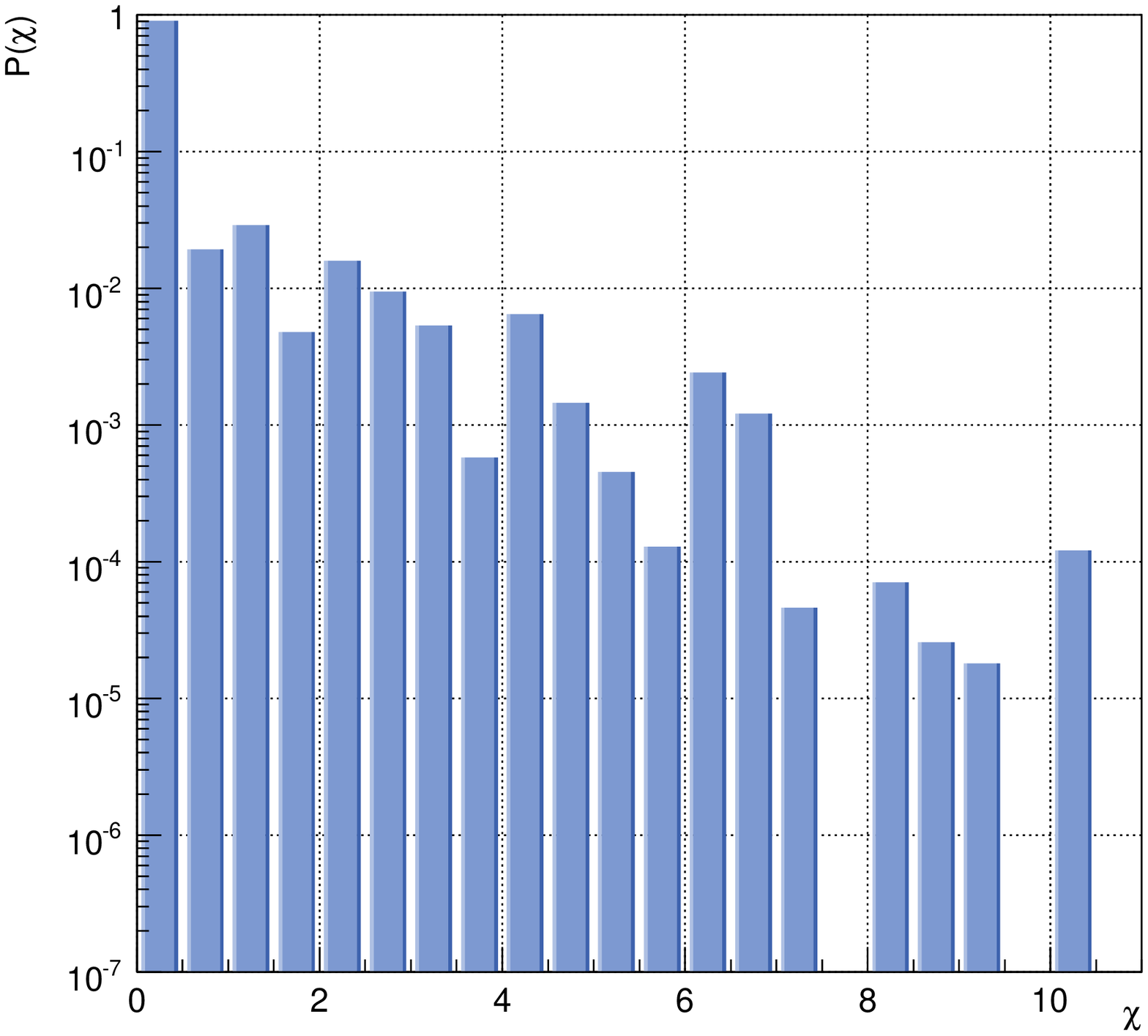}{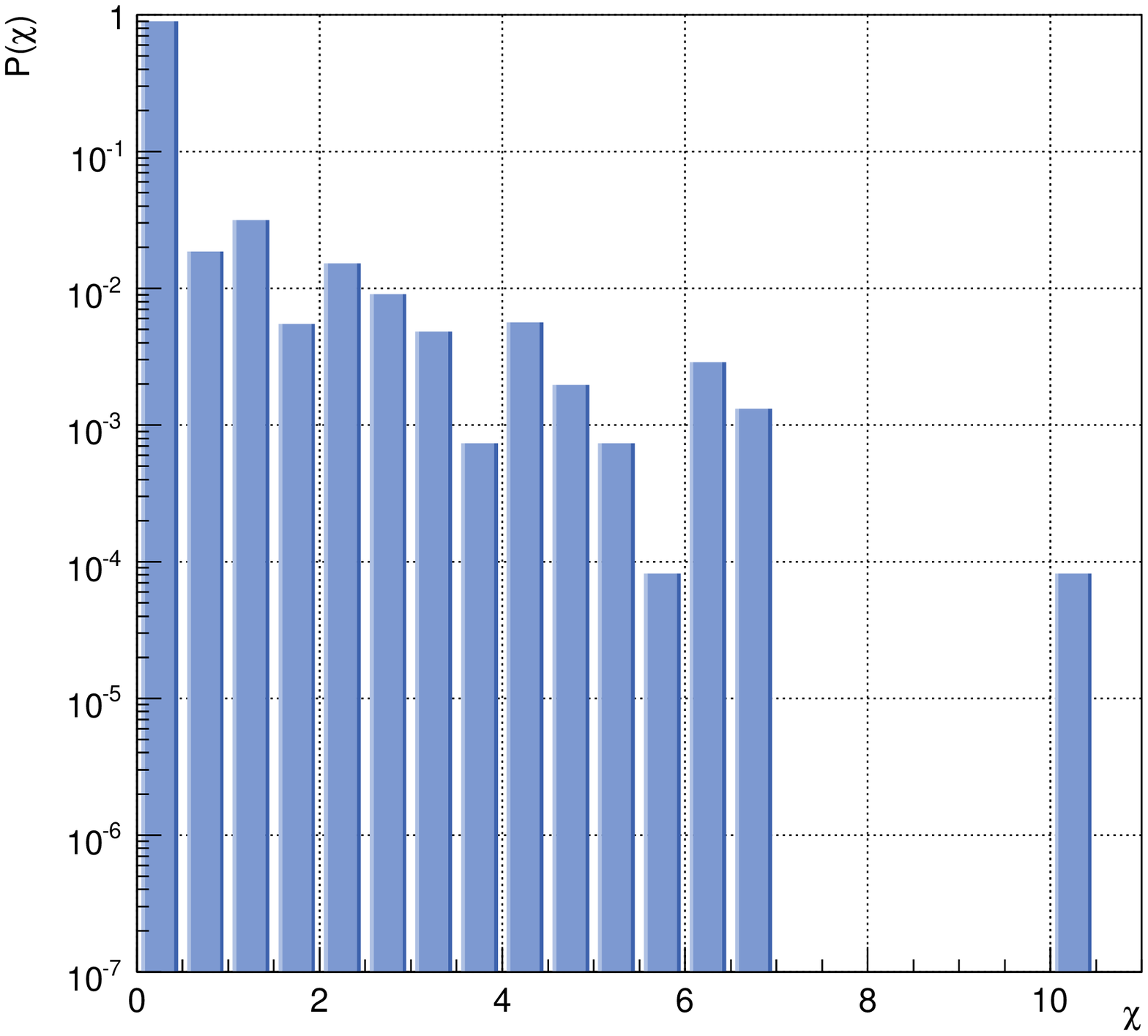}{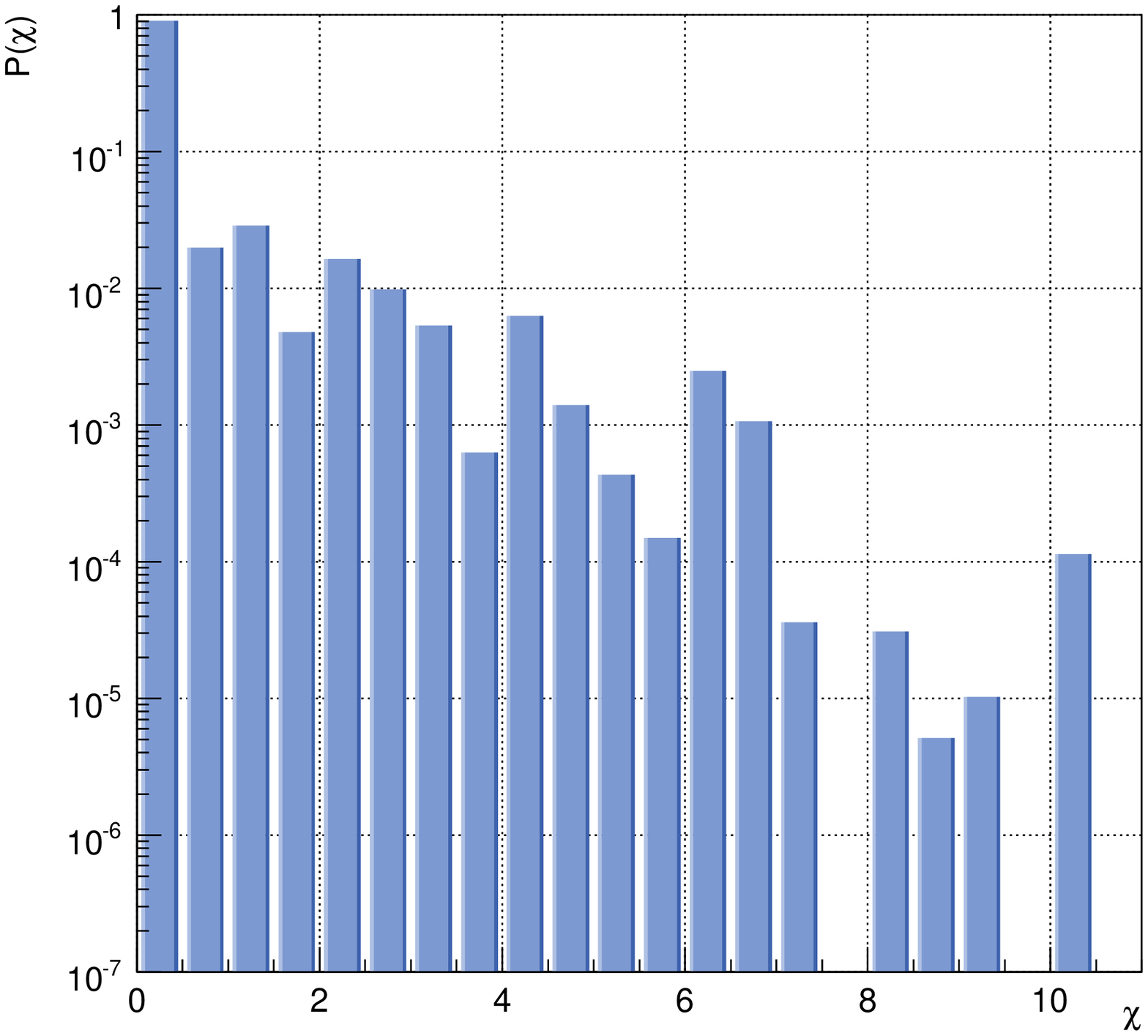}{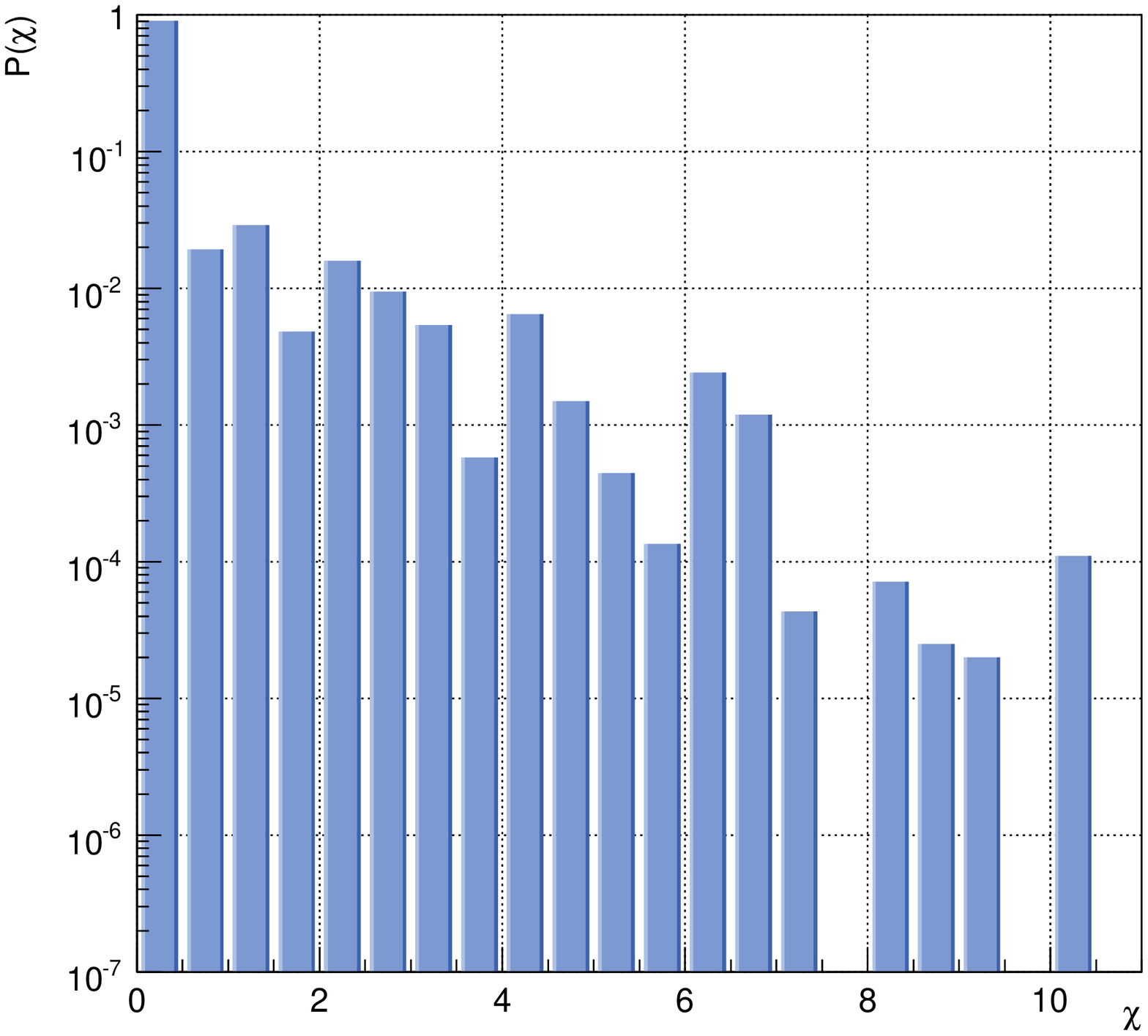}{figchi3}{Frequency
distribution of the mean chirality $\chi$ for models with three stacks of
branes. Shown are the full set of solutions (a) and three sets generated using
randomly chosen subsets of 64~(b), 1024~(c) and 16384~(d) out of all
2097152 possible combinations of exceptional cycles.}

Before considering all random subsets, we have to make sure that
the method can also be trusted in this case, since we are asking a different
question then in the case of gauge factors or rank distributions.
The definition of the mean chirality~\eqref{eqmeanchi} involves a summation
over intersection numbers. These depend very much on the choice of exceptional
cycles, in contrast to the properties of the gauge group, which depend only on
the configuration of bulk cycles.

In Figure~\ref{figchi3} we compare the distribution obtained from the full set
of solutions for models with three stacks of branes, including all
$128^3=2^{21}$ possible choices of exceptional cycles, shown in
Figure~\ref{figchi3_a}, with different random subset--models, shown in
Figures~\ref{figchi3_b},~\ref{figchi3_c} and~\ref{figchi3_d}, which take
64, 1024 and 16384 randomly chosen combinations of exceptional cycles into
account.
Keeping in mind that the plots are logarithmic, one can see that the
qualitative behaviour of the full solution is already captured by the sample
with only 64 randomly chosen exceptional cycles, although we are losing a
good deal of information about models with chirality above $6$.
To obtain a quantitatively satisfying result, it is therefore necessary to
include a bigger subset of cycles. For the highest value of $2^{14}$ random
sets, we get a distribution which differs from the complete result by an
overall error smaller then 1~\textperthousand, comparable to the errors we
found for frequency distributions of gauge group properties.

\twofig{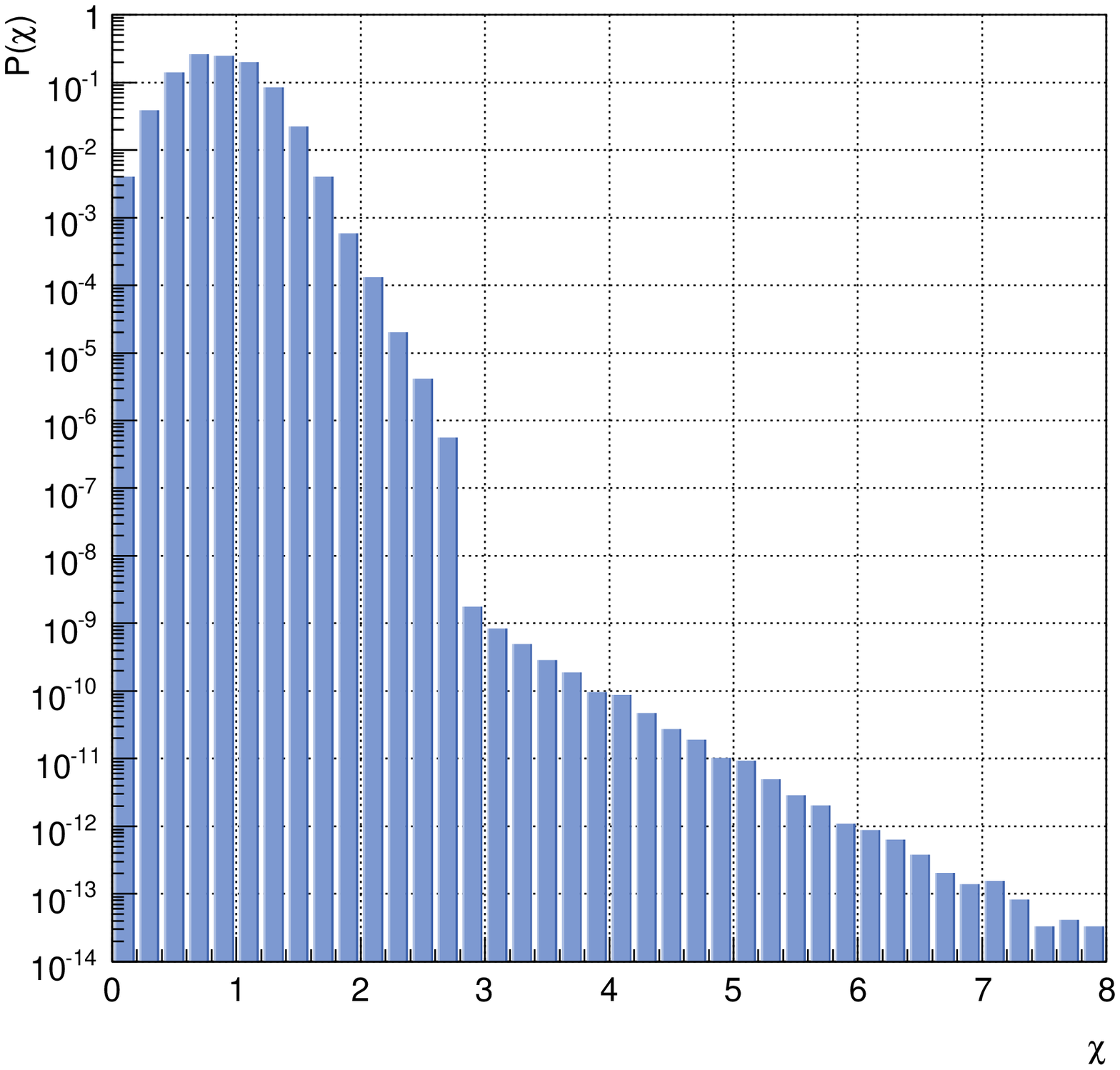}{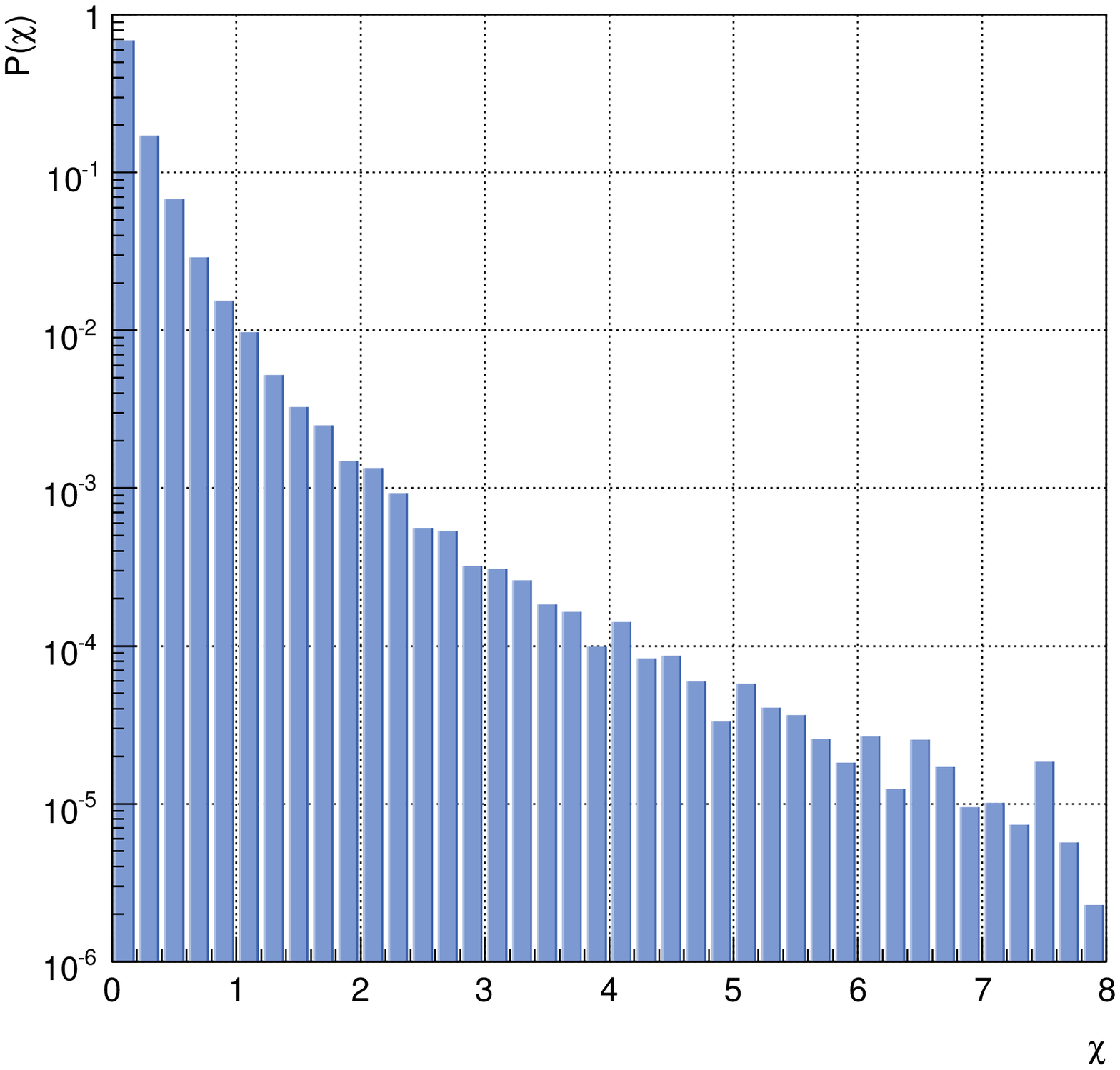}{figchi}{Frequency distribution of models with mean
chirality~$\chi$, as defined in~\protect\eqref{eqmeanchi}, for the present
$\bZ_6$ case~(a) to be compared with the result for $\ZZ$~(b).}

The inclusion of random samples of all possible stack sizes, weighted according
to~\eqref{eqfit}, leads to a frequency distribution as displayed in
Figure~\ref{figchi_a}. Until a value of $\chi\approx 2.8$ the contribution from
models with more than eight stacks dominates. For these models the chirality is
smaller on average, since the $\AAB$ geometry, which allows for solutions with
high chirality is no longer possible.

The distribution is quite different from what has been found for the
$\ZZ$ orbifold. In that case a general scaling behaviour was discovered, that
has been conjectured in~\cite{bghlw04} based on a saddle point
approximation\footnote{An analysis of the mean chirality distribution based on
explicit, computer--generated data can be found in~\cite{gm06}. We will use
this data, which is more accurate then the estimate of~\cite{bghlw04}, to
compare with the present case.}.
The chirality distribution, displayed in Figure~\ref{figchi_b}, scales to a
quite good approximation as $P(\chi) \sim e^{-3\sqrt{\chi}}$.
In the present case the behaviour is different, especially because the
distribution has two parts that scale differently. The first part goes roughly
like $e^{-(\chi^2)}$, while the second part scales like $e^{-\chi}$.

\subsection{Standard model configurations}\label{secsm}
In the following we are going to focus our analysis on a special subclass of
models, namely those which contain the gauge group and the chiral matter
content of the standard model. To be precise, we should speak about the
MSSM here, since all our models are $\cN=1$ supersymmetric.

In order to simplify the analysis we use the term ``standard
model'' in a very broad sense. In this section a standard model refers to
a consistent solution which contains at least the gauge group and the chiral
matter content of the MSSM.
This means that there always exists a hidden sector, containing additional
gauge groups and chiral matter. This is actually not necessarily bad for
phenomenology, since in the end we need a mechanism to break supersymmetry,
which can be nicely accomplished using a mediation through hidden sector
fields.

Concerning so--called ``chiral exotics'', i.e. matter that transforms non
trivially under one of the gauge groups of the standard model, we will
distinguish three cases to make our results comparable with the literature.
Case (i) will have no restrictions on exotic matter at all. In case (ii) we
forbid all exotic matter with the exception of bifundamental representations
of the $SU(2)$ group of the standard model and an additional $U(1)$.
These models are those that have been considered in~\cite{hoot04} and might be
of phenomenological relevance, since the bifundamentals can be interpreted as
supersymmetric Higgs particles. However, since they do not transform under the same
$U(1)$ as the weak doublets, it should be expected that the Yukawa couplings
will be non--standard. In the most restrictive case (iii) we do not allow for
any exotic matter at all.

As has been shown in~\cite{hoot04}, standard model configurations can only
occur if the number of stacks is five or greater. The maximum number of stacks
is nine, since models with more stacks do not support an $SU(3)$ gauge group.
To simplify the analysis we will restrict ourselves to a special type of
embedding of the standard model gauge group and chiral matter, namely the
one introduced in Section~\ref{secsmembed}.

\twofig{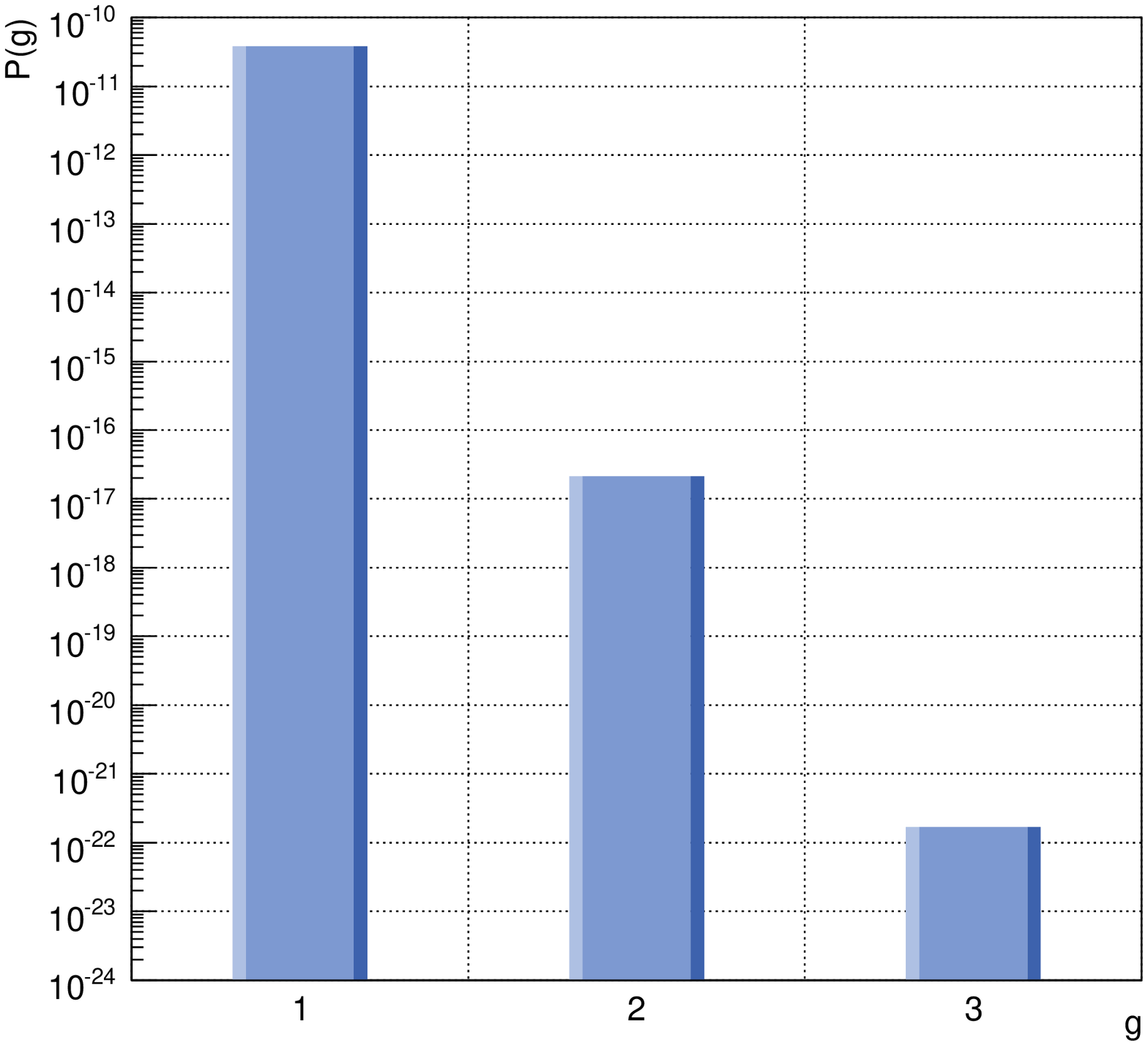}{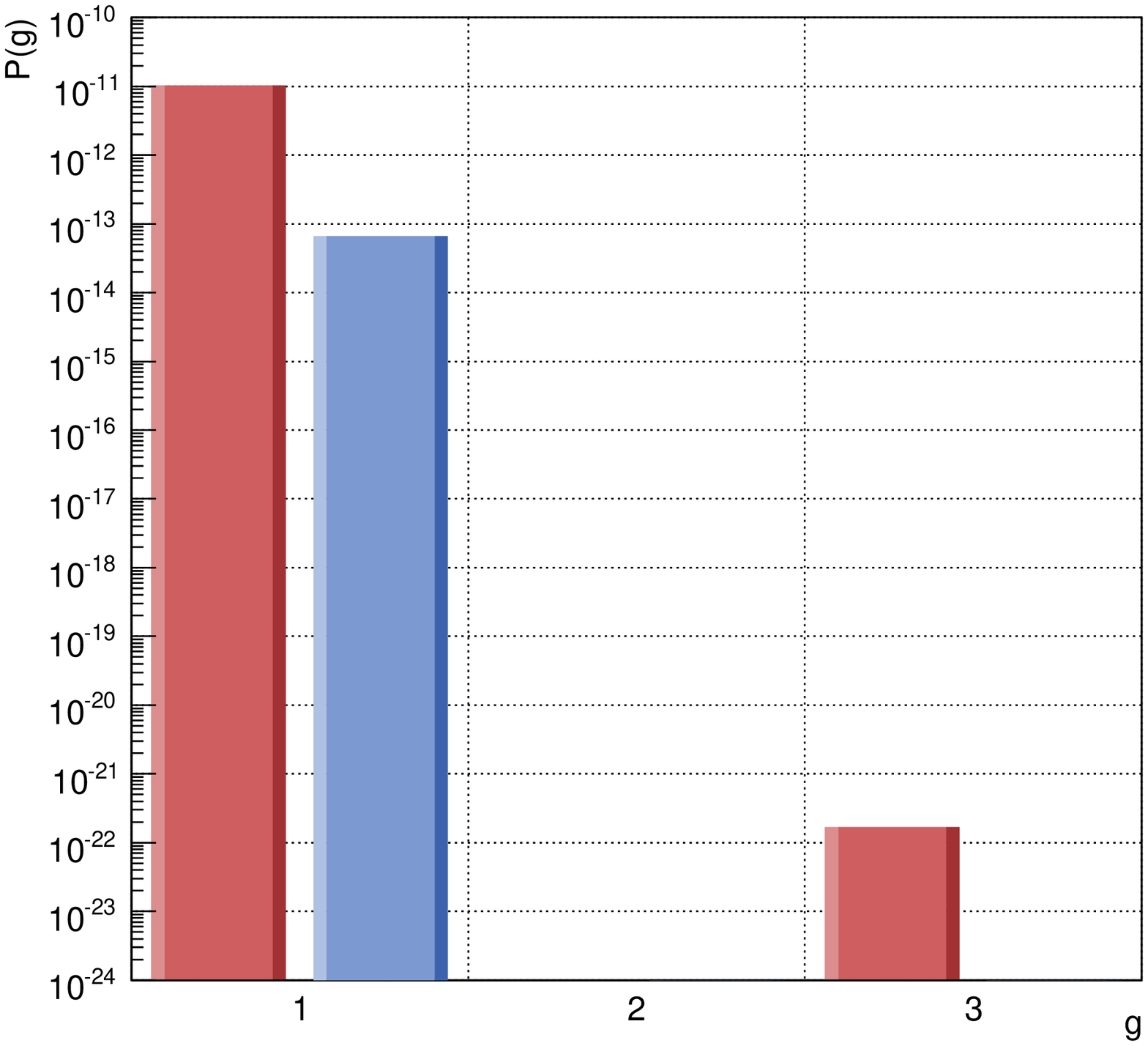}{figgen}{Distributions of the probability to find
models with the gauge group of the standard model and $g$ generations of
the chiral matter content. Figure~(a) shows the results without any
restrictions on chiral exotics, while in~(b) the amount of bifundamental
matter has been restricted either to maximally one pair transforming in the
$SU(2)$ of the standard model (red bars on the left) or to no additional chiral
matter at all (blue bars on the right).}

\subsubsection{Number of generations}
At this point we leave the number of generations of quarks and leptons
as a free parameter. In Figure~\ref{figgen} the frequency distribution of
standard models with different numbers of families is shown.
In Figure~\ref{figgen_a} we allowed all solutions with the gauge group and the
chiral matter content of the standard model, while in Figure~\ref{figgen_b} we
imposed additional constraints to exclude models with chiral exotics, as
outlined above.

Models with more then two generations have only been found in the cases~(i)
and~(ii), which allow for some amount of non--standard matter. In particular,
there are~$\approx 5.7\times 10^6$ solutions with three generations.
These models all contain five stacks of branes and are of
type~(ii), containing one pair of bifundamental matter that transforms in
the $SU(2)$ of the standard model gauge group and the $U(1)$ coming from the
additional fifth brane. The models are of a type similar to 
the ones described in~\cite{hoot04} and contain those as special cases.
We also confirm the statement of that work, that such models can only exist
for the $\AAA$ geometry.

\subsubsection{Spectra}
As already mentioned above, we find the spectrum described in~\cite{hoot04}.
It can be described by the following cycles for the five brane stacks,
\bea{eqspec1}\nonumber
  \pi_a &=& \frac{1}{2}\left(\rho_1+\rho_2 + \e_1-2\te_1-2\e_2+\te_2+\e_5-2\te_5 \right),\\\nonumber
  \pi_b &=& \frac{1}{2}\left(\rho_1+\rho_2-\e_1+2\te_1-2\e_2+\te_2-\e_5+2\te_5\right),\\\nonumber
  \pi_c &=& \frac{1}{2}\left(\rho_1+\rho_2+3\e_2-3\te_2-\e_4-\te_4+\e_5+\te_5\right),\\\nonumber
  \pi_d &=& \frac{1}{2}\left(\rho_1+\rho_2-\e_1+2\te_1+2\e_2-\te_2-\e_5+2\te_5\right),\\\nonumber
  \pi_e &=& \frac{1}{2}\left(\rho_1+\rho_2+3\e_2-3\te_2+\e_4+\te_4-\e_5-\te_5\right).
\eea
The chiral spectrum is given in Table~\ref{tab_smspec1}.

\begin{table}[ht]
\begin{center}%
\begin{equation*}
  \begin{array}{|l|l|c|r|r|r|r|r|r|} \hline
\text{matter}&\text{intersection}&
  SU(3)_a \times SU(2)_b & Q_a & Q_b & Q_c & Q_d & Q_e & Q_Y \\\hline
Q_L & I_{ab'}=-3 & (\overline{\3},\2) & -1 & -1 & 0 & 0 & 0 & \frac{1}{6} \\
U_R& I_{ac}=3 & (\3,\1) & 1 & 0 & -1 & 0 & 0 & -\frac{2}{3} \\
D_R & I_{ac'}=3 & (\3,\1) & 1 & 0 & 1 & 0 & 0 & \frac{1}{3}  \\
L & I_{bd'}=3 & (\1,\2) & 0 & 1 & 0 & 1 & 0 & -\frac{1}{2} \\
E_R & I_{cd}=3 & (\1,\1) & 0 & 0 & 1 & -1 & 0 & 1\\
N_R & I_{cd'}=-3 & (\1,\1) & 0 & 0 & -1 & -1 & 0 &0 \\\hline
& I_{be}=3 & (\1,\2) & 0 & 1 & 0 & 0 & -1 & 0\\
& I_{be'}=3 & (\1,\2) & 0 & 1 & 0 & 0 & 1 & 0\\
\hline
\end{array}
\end{equation*}
\caption{Chiral spectrum of one of the models with gauge group $SU(3)\times SU(2)\times U(1)^5$ that contains the chiral matter spectrum of the standard model. $Q_Y$ is the hypercharge, realised as a combination of all $U(1)$ factors. Intersection numbers that are not listed are zero. Note the explicit appearance of right--handed neutrinos ($N_R$).}
\label{tab_smspec1}
\end{center}
\end{table}

In addition we find several variations of this configuration, all very similar
in structure. The difference between all of these models is only given by the
explicit realisation in terms of wrapping numbers. Furthermore the left--handed
quarks might be realised through $I_{ab}=-3$ and $I_{ab'}$ vanishing.
However, this is only a technical detail and does not change the general setup
of the model, which can therefore be seen as the unique construction to obtain
a standard model spectrum on this particular orbifold.

\subsubsection{Hidden sector}
The hidden sector of the standard models is generically very small.
This is in sharp contrast to the models on the $\ZZ$ orbifold, where we found
quite large hidden sectors with a distribution of gauge groups that turned out
to be almost identical to the distribution in the full set of
models~\cite{gbhlw05}.
The reason for this is that the number of stacks in the
present case is restricted to a maximum of nine, and the tadpole equations limit
the total rank to be lower than or equal to twelve
(cf. Section~\ref{sectotrank}).

If we restrict our attention to the group of models which are most interesting
from a phenomenological point of view, namely the three generation models, we
find that they only occur in configurations with five stacks of branes. In
this case the ``hidden sector'' consists only of one $U(1)$ gauge factor and
in addition we always have chiral matter transforming under this $U(1)$ and
the $SU(2)$ group of the standard model.

%
%
\section{Conclusions and outlook}\label{conclusions}
In this work we have performed a complete analysis of type II intersecting
D--brane models on the $T^6/\bZ_6$ orientifold. We found that there
exist~$3.4\times 10^{28}$ solutions in total, out of which~$5.7\times 10^6$
contain the gauge group and chiral matter content of the standard model.
We therefore obtained a probability of~$1.7\times 10^{-22}$ to find an
MSSM--like vacuum, a number considerably lower then the value of~$10^{-9}$
that has been calculated in the case of $\ZZ$ orientifolds in~\cite{gbhlw05}.

The distribution of gauge groups and chiral matter in the full set of solutions
has been analysed and we compared the results with those from a similar study
of $\ZZ$ models. Similar frequency distributions of single gauge group factors
have been found, but the distribution of the total rank of the gauge group and
of the chiral matter content are quite different. This has been explained by
the fact that the branes considered in this work are actually fractional branes
that wrap not only torus cycles, but generically also exceptional cycles around
fixed points of the orbifold. Since there exists a large number of
possibilities to combine these cycles, the number of solutions is considerably
increased and the statistical distributions are altered significantly compared
to the $\ZZ$--case, in which fractional branes have not been considered.

To obtain the full statistics, a method based on the choice of randomly chosen
subsets of the full solution space has been used. Therefore our results are
not exact, but come with a statistical error, which is however very small and
always below 1\%.

Concerning future directions, it would certainly be very interesting to
compare our results with other string compactifications that use different
setups.
In particular a better comparison with the heterotic
landscape~\cite{di06,le06} and the statistics of M--theory vacua~\cite{acde05}
would be desirable.
Comparing our results with the extensive analysis of Gepner
models~\cite{dhs04a,dhs04b,adks06} would also be very interesting, although
in this case the analysis is complicated by the fact that we are
considering only one particular geometry over a wide range of (untwisted)
moduli here, whereas the analysis in the
works cited above has been done for a very large set of different geometries
at a particular point in moduli space.
Moreover the Gepner model statistic considers only models which resemble the
standard model gauge group. Nevertheless we hope to come back to this issue
in the future.

On a more technical level, our analysis of solutions that resemble properties
of the standard model could be improved. Since we discussed only one possible
embedding there might be more possible realisations with interesting
phenomenological features, although most of the embeddings used in different
contexts will not work due to the fact that the number of symmetric and
antisymmetric representations has to be equal.

Another extension of this work concerns the inclusion of fluxes. In a naive
way this can be done easily by considering a lowered orientifold charge, as
this would generically be the effect of switching on three--form flux.
However, to incorporate the most general NSNS-- and RR--fluxes into an
orientifold setup, it seems very likely that the simple mathematical formalism
used in this article has to be considerably extended.

\acknowledgments
We would like to thank Gabriele Honecker for many valuable discussions and
explanations. We acknowledge interesting conversations with Ralph
Blumenhagen, Tim Dijkstra, Bert Schellekens and Chris White.
The work of F. G. is supported by the Foundation for Fundamental Research of
Matter (FOM) and the National Organisation for Scientific Research (NWO).

%
%
\bibliographystyle{JHEP}
\bibliography{refs_z6}

\providecommand{\href}[2]{#2}\begingroup\raggedright\begin{thebibliography}{10}

\bibitem{lu04}
D.~L{\"u}st, {\it Intersecting brane worlds: {A} path to the standard model?},
  {\em Class. Quant. Grav.} {\bf 21} (2004) S1399--1424,
  [\href{http://xxx.lanl.gov/abs/hep-th/0401156}{{\tt hep-th/0401156}}].

\bibitem{bcls05}
R.~Blumenhagen, M.~Cvetic, P.~Langacker, and G.~Shiu, {\it Toward realistic
  intersecting {D}-brane models},  {\em Ann. Rev. Nucl. Part. Sci.} {\bf 55}
  (2005) 71--139, [\href{http://xxx.lanl.gov/abs/hep-th/0502005}{{\tt
  hep-th/0502005}}].

\bibitem{bkls06}
R.~Blumenhagen, B.~K{\"o}rs, D.~L{\"u}st, and S.~Stieberger, {\it
  Four-dimensional string compactifications with {D}-branes, orientifolds and
  fluxes},  \href{http://xxx.lanl.gov/abs/hep-th/0610327}{{\tt
  hep-th/0610327}}.

\bibitem{lls86}
W.~Lerche, D.~L{\"u}st, and A.~N. Schellekens, {\it Chiral four-dimensional
  heterotic strings from selfdual lattices},  {\em Nucl. Phys.} {\bf B287}
  (1987) 477.

\bibitem{do03}
M.~R. Douglas, {\it The statistics of string / {M} theory vacua},  {\em JHEP}
  {\bf 05} (2003) 046, [\href{http://xxx.lanl.gov/abs/hep-th/0303194}{{\tt
  hep-th/0303194}}].

\bibitem{ku06}
J.~Kumar, {\it A review of distributions on the string landscape},  {\em Int.
  J. Mod. Phys.} {\bf A21} (2006) 3441--3472,
  [\href{http://xxx.lanl.gov/abs/hep-th/0601053}{{\tt hep-th/0601053}}].

\bibitem{doka06}
M.~R. Douglas and S.~Kachru, {\it Flux compactification},
  \href{http://xxx.lanl.gov/abs/hep-th/0610102}{{\tt hep-th/0610102}}.

\bibitem{ddk07}
F.~Denef, M.~R. Douglas, and S.~Kachru, {\it Physics of {S}tring {F}lux
  {C}ompactifications},  \href{http://xxx.lanl.gov/abs/hep-th/0701050}{{\tt
  hep-th/0701050}}.

\bibitem{acdo06}
B.~S. Acharya and M.~R. Douglas, {\it A finite landscape?},
  \href{http://xxx.lanl.gov/abs/hep-th/0606212}{{\tt hep-th/0606212}}.

\bibitem{bghlw04}
R.~Blumenhagen, F.~Gmeiner, G.~Honecker, D.~L{\"u}st, and T.~Weigand, {\it The
  statistics of supersymmetric {D}-brane models},  {\em Nucl. Phys.} {\bf B713}
  (2005) 83--135, [\href{http://xxx.lanl.gov/abs/hep-th/0411173}{{\tt
  hep-th/0411173}}].

\bibitem{gbhlw05}
F.~Gmeiner, R.~Blumenhagen, G.~Honecker, D.~L{\"u}st, and T.~Weigand, {\it One
  in a billion: {M}{S}{S}{M}-like {D}-brane statistics},  {\em JHEP} {\bf 01}
  (2006) 004, [\href{http://xxx.lanl.gov/abs/hep-th/0510170}{{\tt
  hep-th/0510170}}].

\bibitem{gm05}
F.~Gmeiner, {\it Standard model statistics of a type {II} orientifold},  {\em
  Fortsch. Phys.} {\bf 54} (2006) 391--398,
  [\href{http://xxx.lanl.gov/abs/hep-th/0512190}{{\tt hep-th/0512190}}].

\bibitem{gmst06}
F.~Gmeiner and M.~Stein, {\it Statistics of {SU}(5) {D}-brane models on a type
  {II} orientifold},  {\em Phys. Rev.} {\bf D73} (2006) 126008,
  [\href{http://xxx.lanl.gov/abs/hep-th/0603019}{{\tt hep-th/0603019}}].

\bibitem{gm06}
F.~Gmeiner, {\it Gauge sector statistics of intersecting {D}-brane models},
  {\em Fortsch. Phys.} {\bf 55} (2007) 111--160,
  [\href{http://xxx.lanl.gov/abs/hep-th/0608227}{{\tt hep-th/0608227}}].

\bibitem{kuwe05}
J.~Kumar and J.~D. Wells, {\it Surveying standard model flux vacua on
  ${T}^6/\mathbb{Z}_2\times\mathbb{Z}_2$},  {\em JHEP} {\bf 09} (2005) 067,
  [\href{http://xxx.lanl.gov/abs/hep-th/0506252}{{\tt hep-th/0506252}}].

\bibitem{dota06}
M.~R. Douglas and W.~Taylor, {\it The landscape of intersecting brane models},
  {\em JHEP} {\bf 01} (2007) 031,
  [\href{http://xxx.lanl.gov/abs/hep-th/0606109}{{\tt hep-th/0606109}}].

\bibitem{dhs04a}
T.~P.~T. Dijkstra, L.~R. Huiszoon, and A.~N. Schellekens, {\it Chiral
  supersymmetric standard model spectra from orientifolds of {G}epner models},
  {\em Phys. Lett.} {\bf B609} (2005) 408--417,
  [\href{http://xxx.lanl.gov/abs/hep-th/0403196}{{\tt hep-th/0403196}}].

\bibitem{dhs04b}
T.~P.~T. Dijkstra, L.~R. Huiszoon, and A.~N. Schellekens, {\it Supersymmetric
  standard model spectra from {RCFT} orientifolds},  {\em Nucl. Phys.} {\bf
  B710} (2005) 3--57, [\href{http://xxx.lanl.gov/abs/hep-th/0411129}{{\tt
  hep-th/0411129}}].

\bibitem{adks06}
P.~Anastasopoulos, T.~P.~T. Dijkstra, E.~Kiritsis, and A.~N. Schellekens, {\it
  Orientifolds, hypercharge embeddings and the standard model},  {\em Nucl.
  Phys.} {\bf B759} (2006) 83--146,
  [\href{http://xxx.lanl.gov/abs/hep-th/0605226}{{\tt hep-th/0605226}}].

\bibitem{di06}
K.~R. Dienes, {\it Statistics on the heterotic landscape: {G}auge groups and
  cosmological constants of four-dimensional heterotic strings},  {\em Phys.
  Rev.} {\bf D73} (2006) 106010,
  [\href{http://xxx.lanl.gov/abs/hep-th/0602286}{{\tt hep-th/0602286}}].

\bibitem{le06}
O.~Lebedev {\em et~al.}, {\it A mini-landscape of exact {MSSM} spectra in
  heterotic orbifolds},  {\em Phys. Lett.} {\bf B645} (2007) 88--94,
  [\href{http://xxx.lanl.gov/abs/hep-th/0611095}{{\tt hep-th/0611095}}].

\bibitem{hoot04}
G.~Honecker and T.~Ott, {\it Getting just the supersymmetric standard model at
  intersecting branes on the $\mathbb{Z}_6$-orientifold},  {\em Phys. Rev.}
  {\bf D70} (2004) 126010, [\href{http://xxx.lanl.gov/abs/hep-th/0404055}{{\tt
  hep-th/0404055}}].

\bibitem{ddg97}
D.-E. Diaconescu, M.~R. Douglas, and J.~Gomis, {\it Fractional branes and
  wrapped branes},  {\em JHEP} {\bf 02} (1998) 013,
  [\href{http://xxx.lanl.gov/abs/hep-th/9712230}{{\tt hep-th/9712230}}].

\bibitem{digo99}
D.-E. Diaconescu and J.~Gomis, {\it Fractional branes and boundary states in
  orbifold theories},  {\em JHEP} {\bf 10} (2000) 001,
  [\href{http://xxx.lanl.gov/abs/hep-th/9906242}{{\tt hep-th/9906242}}].

\bibitem{dile06}
K.~R. Dienes and M.~Lennek, {\it Fighting the floating correlations:
  Expectations and complications in extracting statistical correlations from
  the string theory landscape},  {\em Phys. Rev.} {\bf D75} (2007) 026008,
  [\href{http://xxx.lanl.gov/abs/hep-th/0610319}{{\tt hep-th/0610319}}].

\bibitem{balo06}
D.~Bailin and A.~Love, {\it Towards the supersymmetric standard model from
  intersecting {D}6-branes on the {Z}'(6) orientifold},  {\em Nucl. Phys.} {\bf
  B755} (2006) 79--111, [\href{http://xxx.lanl.gov/abs/hep-th/0603172}{{\tt
  hep-th/0603172}}].

\bibitem{ghsXX}
F.~Gmeiner and G.~Honecker. Work in progress.

\bibitem{klra00}
M.~Klein and R.~Rabadan, {\it D = 4, {N} = 1 orientifolds with vector
  structure},  {\em Nucl. Phys.} {\bf B596} (2001) 197,
  [\href{http://xxx.lanl.gov/abs/hep-th/0007087}{{\tt hep-th/0007087}}].

\bibitem{bgo02}
R.~Blumenhagen, L.~G{\"o}rlich, and T.~Ott, {\it Supersymmetric intersecting
  branes on the type {I}{I}{A} ${T}^6/\mathbb{Z}_4$ orientifold},  {\em JHEP}
  {\bf 01} (2003) 021, [\href{http://xxx.lanl.gov/abs/hep-th/0211059}{{\tt
  hep-th/0211059}}].

\bibitem{wi82}
E.~Witten, {\it An {S}{U}(2) anomaly},  {\em Phys. Lett.} {\bf B117} (1982)
  324--328.

\bibitem{ur00}
A.~M. Uranga, {\it D-brane probes, {R}{R} tadpole cancellation and {K}-theory
  charge},  {\em Nucl. Phys.} {\bf B598} (2001) 225--246,
  [\href{http://xxx.lanl.gov/abs/hep-th/0011048}{{\tt hep-th/0011048}}].

\bibitem{acde05}
B.~S. Acharya, F.~Denef, and R.~Valandro, {\it Statistics of {M} theory vacua},
   {\em JHEP} {\bf 06} (2005) 056,
  [\href{http://xxx.lanl.gov/abs/hep-th/0502060}{{\tt hep-th/0502060}}].

\end{thebibliography}\endgroup

\end{document}